\documentclass[epj,nopacs%,smallcondensed
]{svjour}
\usepackage[british]{babel}
\def\AnswerYes{y}
\def\MakeArXivLinksActive{y}      %%%%%% Choose whether ArXiv links are active
   \usepackage{grffile} % allows graphics files to have "." in filename
\graphicspath{{figures/}{./}}

    \usepackage[ 
        final,
        breaklinks=true,
        colorlinks=true,           % coloured links 
        pdfborder={0 0 1},    % coloured box about link
        citecolor={blue},linkcolor={blue},urlcolor={red},
        citebordercolor={1 0 0},linkbordercolor={0 0 1},urlbordercolor={1 0 0},
        pdfpagemode=UseOutlines,% whether Acrobat opens with
        bookmarks=true,bookmarksopenlevel=4% write .pdf table of contents.
        ]{hyperref}         % Hyperlinks for PDF versions

\ifx\MakeArXivLinksActive\AnswerYes
   \usepackage{xparse}
   \NewDocumentCommand{\arxiv} %
   {r [: u{ [} u{]]} }{[\href{http://arxiv.org/abs/#2}{arXiv:#2}~[#3]]}
   \NewDocumentCommand{\arxivold} {r[]}{[\href{http://arxiv.org/abs/#1}{#1}]}
   \NewDocumentCommand{\arXiv} %
   {r [: u{ [} u{]]} }{[\href{http://arxiv.org/abs/#2}{arXiv:#2}~[#3]]}
   \NewDocumentCommand{\arXivold} {r[]}{[\href{http://arxiv.org/abs/#1}{#1}]}
\else
   \newcommand{\arxiv}[1][]{[#1]}
   \newcommand{\arxivold}[1][]{[#1]}
   \newcommand{\arXiv}[1][]{[#1]}
   \newcommand{\arXivold}[1][]{[#1]}
\fi
\usepackage{slashed}
\usepackage{multirow}                   %%% tabs over multiple rows
\usepackage[numbers,sort&compress]{natbib}%%% collapses [1,2,3] -> [1-3] etc
\usepackage{graphicx} % for figures
\usepackage{bm}       % bold math 
\usepackage{amssymb}
\usepackage{amsmath}
\usepackage{xspace}                    %%% correct spacing self-defined macros

\newcommand{\ii}{\mathrm{i}}
\newcommand{\dd}{\mathrm{d}}

\newcommand{\half}{\frac{1}{2}}

\newcommand{\mpi}{\ensuremath{m_\pi}}     % pion mass
\newcommand{\MeV}{\ensuremath{\mathrm{MeV}}}

\newcommand{\ChiEFT}{$\chi$EFT\xspace}

\newcommand{\kv}{\vec{k}} 
\newcommand{\qv}{\vec{q}}

\newcommand{\ptyp}{p_\text{typ}}
\newcommand{\LambdaEFT}{\overline{\Lambda}_\text{EFT}}
\newcommand{\LambdaNoPion}{\overline{\Lambda}_{\slashed{\pi}}}
\newcommand{\EFTNoPion}{EFT($\slashed{\pi}$)\xspace}
\newcommand{\LambdaChi}{\overline{\Lambda}_{\chi}}

\newcommand{\NXLO}[1]{N\ensuremath{{}^{#1}}LO\xspace}

\newcommand{\hq}{\hspace{0.5ex}}
\newcommand{\absatz}{\vspace{2ex}\noindent}

\newcommand{\calA}{\mathcal{A}}
\newcommand{\calC}{\mathcal{C}}

\newcommand{\calO}{\mathcal{O}}

\newcommand{\wave}[3]{\ensuremath{{}^{#1}\mathrm{#2}_{#3}}}

\newcommand{\eg}{\emph{e.g.\xspace}} 
\newcommand{\ie}{\emph{i.e.\xspace}}
\newcommand{\cf}{\emph{cf.\xspace}}

\begin{document}
\title{A Consistency Test of EFT Power Countings from Residual Cutoff
  Dependence}
% \subtitle{Do you have a subtitle?\\ If so, write it here}
\author{Harald W.~Grie{\ss}hammer%\inst{1}\inst{2}% etc
\thanks{\emph{Present address: 1; E-mail: hgrie@gwu.edu}}
%Insert the address here if needed}%
}                     % Do not remove
\authorrunning{H.~W.~Grie{\ss}hammer}
\institute{Institute for Nuclear Studies, Department of Physics, George
  Washington University, Washington DC 20052, USA \and Department of Physics,
  Duke University, Box 90305, Durham NC 27708, USA}
\date{Received: date / Revised version: date}
\abstract{I summarise a method to quantitatively assess the consistency of
  power-counting proposals in Effective Field Theories which are
  non-perturbative at leading order. It uses the fact that the Renormalisation
  Group evolution of an observable predicts the functional form of its
  residual cutoff dependence on the breakdown scale of an Effective Field
  Theory (EFT), on the low-momentum scales, and on the order of the
  calculation.  Passing this test is a necessary but not sufficient
  consistency criterion for a suggested power counting whose exact nature is
  disputed.  For example, in Chiral Effective Field Theory (\ChiEFT) with more
  than one nucleon, a lack of universally accepted analytic solutions
  obfuscates the relation between convergence pattern and numerical results,
  and led to proposals which predict different numbers of Low Energy
  Coefficients (LECs) at the same chiral order, and at times even predicts a
  different ordering long-range contributions. The method may provide an
  independent check whether an observable is properly renormalised at a given
  order, and allows one to estimate both the breakdown scale and the
  momentum-dependent order-by-order convergence pattern of an EFT. Conversely,
  it may help identify those LECs (and long-range pieces) which produce
  renormalised observables at a given order. I also discuss its underlying
  assumptions and relation to the Wilsonian Renormalisation Group Equation;
  useful choices for observables and cutoffs; the momentum window in which the
  test likely provides best signals; its dependence on the values and forms of
  cutoffs as well as on the EFT parameters; the impact of fitting LECs to data
  in different or the same channel; and caveats as well as limitations. Since
  the test is designed to minimise the use of data, it allows one to
  quantitatively falsify if the EFT has been renormalised consistently. This
  complements other tests which quantify how an EFT compares to
  experiment. Its application in particular to the \wave{3}{P}{0} and
  \wave{3}{P}{2}-\wave{3}{F}{2} partial waves of $\mathrm{NN}$ scattering in
  \ChiEFT may elucidate persistent power-counting issues.
%
% \PACS{
%       {PACS-key}{discribing text of that key}   \and
%       {PACS-key}{discribing text of that key}
%      } % end of PACS codes
} %end of abstract
\maketitle
%%%%%%%%%%%%%%%%%%%%%%%%%%%%%%%%%%%%%%%%%%%%%%%%%%%%%%%%%%%%%%%%%%%
%%%%%%%%%%%%%%%%%%%%%%%%%%%%%%%%%%%%%%%%%%%%%%%%%%%%%%%%%%%%%%%%%%%
\section{Motivation: Serious Theorists Have Error Bars}

\subsection{Introduction}

That our understanding of natural phenomena is based on concrete, falsifiable
predictions is deeply ingrained in the scientific method.
It is insufficient to compare numbers; one also must judge their reliability.
And since we do not trust experiments without error bars, why should it be
acceptable for a theorist to not assess uncertainties in a calculation,
\emph{before} a closer look at the data to be explained?  Simply stating that
this is ``difficult'' is certainly no sufficient excuse, especially after the
recent surge of articles which offer statistically meaningful methods to
ascertain and interpret theory errors; see \emph{e.g.}~\cite{editorial, JPhysG,
  JPhysG2}. The prospect of a reproducible, objective, quantitative estimate
of theoretical uncertainties lies thus at the heart not only of any Effective
Field Theory (EFT). But EFTs claim to possess well-defined schemes to find
just such estimates. It is therefore befitting to explore how the validity of
such prescriptions can be gauged.

More than two decades ago, Lepage discussed in a highly influential lecture
methods to quantify convergence to data~\cite{Lepage:1997cs}, no doubt based
on standard lore in Computational Physics (see
\emph{e.g.}~Ref.~\cite{Landau}). In contradistinction, the test presented in
the following hopes to quantify the internal consistency of an EFT and takes
\emph{minimal resort} to experimental information. In an ideal world,
theorists would perform ``double-blind'' calculations in which theoretical
uncertainties are assessed under the pretence that no or only very limited
data is available. Such ``post-dictions'' are of course predictions when
information is indeed experimentally unknown or hard to access, or when data
consistency must be checked. But even when not, they contribute to the
discourse and hopefully increase the community's confidence in the method
used. They thus form an important sociological aspect of the Scientific
Method.

\absatz The presentation is organised as follows. The remainder of this
Introduction aims to motivate why a reliable scheme to quantify theory
uncertainties is imperative but not quite straightforward in EFTs with
unnatural shallow scales, and lists schemes discussed in the
literature. Section~\ref{sec:test} provides a first description of the
proposed test, on a rather abstract level. An example in
sect.~\ref{sec:application} provides a concrete experience of the
method. After this illustration, the final section addresses a number of what
may at first reading be perceived as details. Most of these are
cross-referenced in sect.~\ref{sec:test}, so that the interested reader can
quickly find further information if desired. These topics include: extensions,
assumptions, choices, and limitations of the method; a discussion of which
observables and kinematics are likely to provide clear signals; and
miscellaneous notes; followed by concluding remarks. This organisation has the
advantage to not overburden the initial presentation in sect.~\ref{sec:test}
with a large numbers of qualifiers and footnotes which disrupt the flow.

On a historical note, the origin of these remarks goes back to publications in
2003 and 2005~\cite{Bedaque:2002yg, improve3body}, and to lectures at the 2008
US National Nuclear Physics Summer School~\cite{NNPSS}. When the issue was
revisited at two % more recent
workshops~\cite{Saclay, Benasque}, its conclusions were generally perceived as
not immediately straightforward or widely known. Input on some aspects was
also provided for two recent publications~\cite{Furnstahl:2014xsa,
  Epelbaum:2014efa}. Dai \emph{et~al.}~explored it in
$\overline{\mathrm{N}}\mathrm{N}$ scattering~\cite{Dai:2017ont}. It seems
therefore fitting to present an expanded Technical Note, updating, expanding
and -- where necessary -- correcting a proceeding from
2015~\cite{Griesshammer:2015osb}.

%%%%%%%%%%%%%%%%%%%%%%%%%%%%%%%%%%%%%%%%%%%%%%%%%%%%%%%%
\subsection{Rationale: Consistency Issues in EFTs with Shallow Bound States}

Effective Field Theories take advantage of a separation of scales to expand
interactions and observables in a dimension-less quantity which for the sake
of this presentation is denoted by
\begin{equation}
\label{eq:Q}
  Q=\frac{\text{typical low momenta } k,\ptyp}{\text{breakdown scale } \LambdaEFT}<1\;\;.
\end{equation}
The numerator contains the relative momentum $k$ between scattering particles,
and other intrinsic low scales which are summarily denoted by $\ptyp$. At the
breakdown scale $\LambdaEFT$, new dynamical degrees of freedom enter which are
not explicitly accounted for by the EFT but whose effects at these short
distances are simplified into Low-Energy Coefficients (LECs).

Consider, for example, Chiral Effective Field Theory (\ChiEFT), the extension
of (purely mesonic) Chiral Perturbation Theory to the one- and few-nucleon
system; see \emph{e.g.}~\cite{Hammer:2019poc} for a recent review: $\ptyp$
then includes the pion mass and the inverse scattering lengths of the
$\mathrm{NN}$ system: $\gamma_t\approx45\;\MeV$ in the \wave{3}{S}{1} channel
and $\gamma_s\approx-8\;\MeV$ in the \wave{1}{S}{0} channel are at leading
order given by the inverse scattering lengths of these channels,
$\gamma_{t,s}\approx1/a_{t,s}$. Its breakdown scale
$\LambdaChi\approx[700\dots1000]\;\MeV$ is consistent with the masses of the
$\omega$ and $\rho$ as the next-lightest exchange mesons, and with the chiral
symmetry breaking scale -- if the $\Delta(1232)$ as lowest nucleonic resonance
is included as dynamical degree of freedom. In a theory without it, the
breakdown scale shrinks to
$\LambdaChi(\slashed{\Delta})\approx300\;\MeV$. Another example is ``pion-less
EFT'' (\EFTNoPion), which in turn is the low-energy version of \ChiEFT because
the pion itself is integrated out, so that
$\LambdaNoPion\approx\mpi\gg \ptyp\sim\gamma_{t,s}$. With such a parameter,
all interactions and contributions to any observable are determined by
expanding them in $Q$ and estimating their relative strengths by Na\"ive
Dimensional Analysis~\cite{NDA,NDA2,Weinberg:1989dx, Georgi:1992dw,
  Griesshammer:2005ga}.  When all interactions are perturbative, as in the
mesonic and one-baryon sectors, this amounts to not much more than counting
powers of $k$ and $\ptyp$ -- hence the name \emph{power counting (PC)} scheme.

The situation is more complicated when some interactions must be treated
non-perturbatively at leading order because of shallow real or virtual bound
states with scales $\ptyp\ll\LambdaEFT$ in the EFT's range of validity. In
$\mathrm{NN}$ scattering, all terms in the leading-order (LO)
Lippmann-Schwinger equation, including the potential, must be of the same
order when all nucleons are close to their non-relativistic mass-shell. If
that were not the case, one term could be treated as perturbation of the
others and there would be no shallow bound-state. Thus, a shallow bound state
necessitates resumming at least some particular interaction or interactions,
\emph{i.e.}~an infinite number of interaction points at LO. This, in turn,
imposes a consistency condition on that interaction and on the amplitude. In a
non-relativistic theory, both the amplitude $T_\mathrm{NN}$ and potential
$V_\mathrm{NN}$ must be of order $Q^{-1}$.

In order to show this, consider the intermediate-state $\mathrm{NN}$
propagator (correlated Green's function)
$G_{\mathrm{NN}}\propto\frac{1}{\qv^2-\kv^2}$ in time-ordered perturbation
theory, where $\qv$ is the relative momentum of one nucleon in the
intermediate state, and $\kv$ is the scattering momentum. In a
field-theoretical approach, $G_{\mathrm{NN}}$ emerges after one picks the
nucleon-pole in the (non-relativistic) energy-integration,
$q_0\sim\frac{q^2}{2M}$ (``potential'' r\'egime; see
\emph{e.g.}~\cite{Beneke:1997zp, Griesshammer:1997wz}). In either case, the
integral is dominated by momenta $k\sim q$.

Denoting $T_\mathrm{NN}$ by an ellipse and  $V_\mathrm{NN}$ by a rectangle,
one arrives at the semi-graphical equation:
\begin{equation}
  \label{eq:consistency}
  \begin{split}
    \includegraphics[clip=,width=\columnwidth]{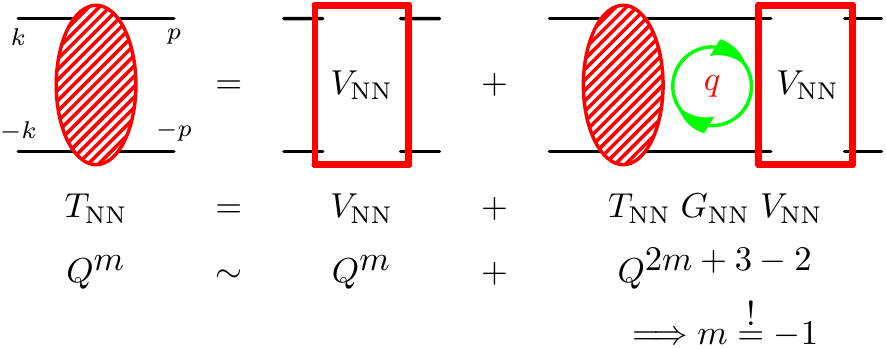}
  \end{split}
\end{equation}
Therefore, a shallow real or virtual bound state mandates
$T_{\mathrm{NN}}\sim V_{\mathrm{NN}}\sim Q^{-1}$, unless one is willing to
resort to \emph{ad-hoc} fine-tuning between differing contributions.

This leads to a surprising take on the one-pion exchange. It appears to scale
as
$(\vec{\sigma}_1\cdot\vec{q})(\vec{\sigma}_2\cdot\vec{q})/
(\vec{q}^2+\mpi^2)\sim Q^0$
when one counts only explicit low-momentum scales, but must be of order
$Q^{-1}$ if its iteration is to be mandated in bound-state dynamics
$q_0\sim\frac{q^2}{2M}$.  For $q_0\sim q$, on the other hand, the nucleon
becomes static, $\mathrm{NN}$ rescattering is suppressed, and one enters the
soft and ultra-soft r\'egimes of Non-Relativistic QED/QCD; see
\emph{e.g.}~Refs.~\cite{Beneke:1997zp, Griesshammer:1997wz} and references
both in and to these.

In a straightforward extension, amplitude and interaction between $n$ nucleons
which accommodate non-pertur\-bative features at LO scale as
\begin{equation}
  \label{eq:consistency-n}
  T_{n\mathrm{N}}\sim V_{n\mathrm{N}}\sim Q^{1-n}\;\;.
\end{equation} 
Since both the interactions and the LECs themselves carry inverse powers of
$Q$, finding their importance by just counting momenta is therefore clearly
insufficient.

The resulting scaling for $T_{n\mathrm{N}}$ only assumes the existence (but
not particular numerical value) of unnaturally small scales, irrespective of
the form of the interaction. It does not reveal \emph{which} terms constitute
the LO potential or how the ``unnatural'' shallow scale emerges; only how
those terms must be power-counted.

These arguments imply that a choice of EFTs exists, all of which have the same
symmetries and degrees of freedom but differ in their power counting. These
include the popularly used version with shallow $\mathrm{NN}$ bound states and
the one-pion exchange at leading order~\cite{Hammer:2019poc}; a ``KSW''
version in which the one-pion exchange scales indeed as $Q^0$ and thus is
\emph{not} needed to bind the shallow bound states at LO but enters at NLO,
with analytic results in the $\mathrm{NN}$ system that pass every test of
self-consistency but limit the radius of convergence to
$\overline{\Lambda}_\mathrm{KSW}\lesssim200\;\MeV$ or so~\cite{Kaplan:1998tg,
  Kaplan:1998we, Fleming:1999ee}; and a version in which low-energy Nuclear
Physics has no shallow bound states at all ($\gamma_{t,s}\sim\LambdaEFT$), so
that the one-pion exchange is perturbative. Nature's provision of shallow
bound states rules out the latter.

In $\mathrm{NN}$ scattering, a main issue appears to be an intrinsic scale
$\Lambda_\mathrm{NN}=16\pi f_\pi^2/(g_A^2 M)\approx300\;\MeV$, where
$f_\pi\approx92\;\MeV$ is the pion decay constant and $g_A\approx1.27$. It
sets the strength of the one-pion exchange potential~\cite{Kaplan:1998we,
  Barford:2002je} and lies right between the typical low scale $\mpi$ and the
expected breakdown scale $\LambdaChi$, possibly providing a less-than-perfect
separation of scales and convergence.

This behaviour of LO amplitudes and interactions in the presence of shallow
bound states has long been recognised in pionless EFT (\EFTNoPion) and its
variants (like Halo-EFT and EFT of point-like interactions), where the scaling
of operators and the $\beta$ functions of couplings be\-tween up to $3$
nucleons are well-established~\cite{Bedaque:2002mn, Platter:2009gz, Hammer:2019poc}. For example, analytic results in well-controlled limits show
one momentum-independent $3\mathrm{N}$ operator at LO. Likewise,
Non-Relativistic QED and QCD count the Coulomb potential as $Q^{-1}$ to allow
its resummation.

The situation in \ChiEFT for two and more nucleons is less obvious. In order
to circumvent the apparent paradox between having to resum in order to get a
shallow bound state on the one hand, and the momentum scaling of the one-pion
exchange potential on the other hand, Weinberg pragmatically suggested to
still count LECs and the one-pion exchange as $Q^0$, but to apply the
perturbative counting of momenta not to amplitudes but to the few-nucleon
potential, which is then iterated to produce shallow bound states.  How this
translates into a PC of observables is under dispute. It has also been
demonstrated that Weinberg's pragmatic proposal predicts an incorrect scaling
of short-distance singularities with $\mpi$~\cite{Beane:2001bc}; and that it
permits no unique convergence for $\Lambda\gtrsim\LambdaEFT$, including in the
limit $\Lambda\to\infty$~\cite{Nogga:2005hy}; see also U.~van Kolck's
contribution to this volume~\cite{vankolck}. Further disagreement persists
about the interpretation of approximate solutions (large off-shell momenta,
semi-classical limit, etc.), and about more technical problems (deeply bound
states etc.). In addition, a cutoff/regulator $\Lambda$ becomes numerically
necessary. It is conceptually quite different from the breakdown scale
$\LambdaEFT$, albeit the two symbols are similar. It cannot be much smaller
than the breakdown scale in order not to ``cut out'' physical, low-resolution
momenta in loops. But even how far $\Lambda$ should be varied is under
dispute: Is any value $\Lambda\gtrsim\LambdaEFT$ legitimate, including
$\Lambda\to\infty$ (see \eg~ref.~\cite{Nogga:2005hy} and references to it); or
should the range be constrained to $\Lambda\approx\LambdaEFT$ (see
\eg~ref.~\cite{Epelbaum:2006pt} and references to it)?

In any case, it should be clear that no particular value of any cutoff
(including infinity) is preferred to any other, as long as
$\Lambda\gtrsim\LambdaEFT$. For any cutoff, one truncates some non-physical
high-momentum/short-distance modes whose Physics is not represented in detail,
just because an EFT's effective degrees of freedom are ineffective for loop
momenta $q\gtrsim\LambdaEFT$.

It is thus no surprise that in addition to Weinberg's pragmatic proposal,
three active PC proposals emerged in \ChiEFT, all with the same degrees of
freedoms: nucleons and pions only~\cite{Weinberg, Birse:2005um, Birse:2009my,
  %PavonValderrama:2005wv, PavonValderrama:2005uj, 
  Valderrama:2009ei, Valderrama:2011mv,Valderrama:2019lhj, Long:2011xw,
  Long:2012ve}. Table~\ref{tab:pc} lists their predictions for the order at
which a LEC or long-range contribution enters in the lower $\mathrm{NN}$
partial waves. While there is agreement that the long-range part is dominated
by one-pion-exchange, ordering of its two-pion exchange is under dispute. More
importantly, each finds a different number of LECs at any given order -- and
each claims consistency.  Not all can be right, though.
Articles, panels and sessions at Chiral Dynamics and other conferences as well
as dedicated workshops led to little consensus (see \emph{e.g.}~Daniel
Phillips' even-handed account at Chiral Dynamics 2012~\cite{Phillips:2013fia}
and van Kolck's more recent review~\cite{vanKolck:2020llt}); some additional
notable contributions include Refs.~\cite{Kaplan:1996xu, Kaplan:1998we,
  Beane:2001bc, Nogga:2005hy, Epelbaum:2006pt, Beane:2008bt, Epelbaum:2009sd,
  Valderrama:2016koj, Hammer:2019poc}.

This is not just stamp-collecting or a philosophical question which
potentially exposes the soft underbelly of \ChiEFT and the credibility of its
error assessments, but which is ``otherwise'' of little practical
consequence. A central EFT promise is that it encodes the unresolved
short-distance information at a given accuracy into not just some, but the
\emph{smallest-possible} number of independent LECs under given
symmetries. For example, the PC proposals of $\mathrm{NN}$ \ChiEFT differ most
for attractive triplet waves: the \wave{3}{P}{2}-\wave{3}{F}{2} system at
order $Q^0$ has no LEC parameter~\cite{Weinberg} -- or $3$ of similar
size~\cite{Birse:2005um, Birse:2009my} -- or $3$, but with different
weights~\cite{%PavonValderrama:2005wv, PavonValderrama:2005uj,
  Valderrama:2009ei, Valderrama:2011mv,Valderrama:2019lhj} -- or
$1$\cite{Long:2011xw, Long:2012ve}. To bring it to a boil: If all proposals
are renormalised and fit $\mathrm{NN}$ data with the same $\chi^2$, the one
with the least number of parameters wins.

For the sake of this article, I am agnostic about this dispute. Rather, I
propose to test if a predicted convergence pattern is reflected in the
answers, \emph{i.e.}~if a proposed power counting is
consistent. % As alluded to at
% the beginning, a consistent power-counting is central to any EFT: only when
% all contributions can be ordered by relative size before calculating
% specific terms, can residual uncertainties be estimated reliably. So, how to
% guarantee that the assumptions of a power counting bear out in the final
% results?

%\begin{strip}
%%%%%%%%%%%%%%%%%%%%%%%%%%%%%%%%%
\begin{table*}[!t]
%\begin{minipage}{\textwidth}
%\onecolumn
\centering
\scriptsize%\tiny
\newcommand{\makerow}[5]{#1&#2&#5&#4&#3} 
\begin{tabular}
%{|l||p{0.18\linewidth}|p{0.187\linewidth}|p{0.215\linewidth}|p{0.205\linewidth}|}
{|l||p{0.2\linewidth}|p{0.2\linewidth}|p{0.2\linewidth}|p{0.2\linewidth}|}

    \hline
    \makerow{\rule[-1.2ex]{0pt}{4ex}\textbf{order}}
    {\textbf{Weinberg} (modified)~\cite{Weinberg}}
    {\textbf{Long/Yang}~2012 \cite{Long:2011xw, Long:2012ve}}
    {\textbf{Pavon et al.}~2006 \cite{% PavonValderrama:2005wv,
        % PavonValderrama:2005uj, 
  Valderrama:2009ei, Valderrama:2011mv,Valderrama:2019lhj}}
    {\textbf{Birse} 2005 \cite{Birse:2005um, Birse:2009my}}
    \\\hline\hline
    \makerow{\rule[-1.2ex]{0pt}{4ex}{$Q^{-1}$}}{ 
      LO of \wave{1}{S}{0}, \wave{3}{S}{1}, OPE}{LO of \wave{1}{S}{0}, \wave{3}{S}{1}, OPE, \wave{3}{P}{0,2}}{
      LO of \wave{1}{S}{0}, \wave{3}{S}{1}, OPE, \wave{3}{P}{0,2}, \wave{3}{D}{2}}{
      LO of \wave{1}{S}{0}, \wave{3}{S}{1}, OPE, \wave{3}{D}{1}, \wave{3}{SD}{1}
    }\\\hline
    \makerow{\rule[-1.2ex]{0pt}{5ex}$Q^{-\half}$}{none}{none}{
      LO of \wave{3}{SD}{1}, \wave{3}{D}{1}, \wave{3}{PF}{2}, \wave{3}{F}{2}}{
      LO of \wave{3}{P}{0,1,2}, \wave{3}{PF}{2}, \wave{3}{F}{2}, \wave{3}{D}{2}
    }\\\hline% \hline
    \makerow{\rule[-1.2ex]{0pt}{4ex}$Q^{0}$}{none}{
      NLO of \wave{1}{S}{0}}{NLO of \wave{1}{S}{0}}{NLO of \wave{1}{S}{0}
    }\\\hline
    \makerow{\rule[-1.2ex]{0pt}{5ex}$Q^{\half}$}{none }{none}{none}{ NLO of \wave{3}{S}{1}, \wave{3}{D}{1}, \wave{3}{SD}{1} }\\
    \hline
    \makerow{\rule[-1.2ex]{0pt}{4ex}\multirow{2}{*}{$Q^1$}}{LO of \wave{3}{SD}{1},\wave{1}{P}{1}, \wave{3}{P}{0,1,2},
        TPE;
      NLO of \wave{1}{S}{0}, \wave{3}{S}{1}}{LO of \wave{3}{SD}{1},\wave{1}{P}{1}, \wave{3}{P}{1},
      \wave{3}{PF}{2}, TPE;
      NLO of  \wave{3}{S}{1}, \wave{3}{P}{0}, \wave{3}{P}{2};
      \NXLO{2} of \wave{1}{S}{0}
    }{\multirow{2}{*}{none}}{\multirow{2}{*}{none}}\\\hline
    \makerow{\rule[-1.2ex]{0pt}{5ex}$Q^\frac{3}{2}$}{none}{none
    }{none}{NLO of \wave{3}{D}{2}, \wave{3}{P}{0,1,2}, \wave{3}{PF}{2}, \wave{3}{F}{2}
    }\\\hline
    \makerow{\rule[-1.2ex]{0pt}{4ex}\multirow{2}{*}{$Q^2$}}{\multirow{2}{*}{NLO of TPE}}{ \multirow{2}{*}{NLO of TPE; \NXLO{3} of \wave{1}{S}{0}}}{
      LO of TPE, \wave{1,3}{P}{1};
      NLO of \wave{3}{S}{1}, \wave{3}{D}{1,2}, \wave{3}{SD}{1}, \wave{3}{P}{0,2},
      \wave{3}{PF}{2}; \NXLO{2} of
      \wave{1}{S}{0}}{
      LO of TPE, \wave{1}{P}{1}; NLO of OPE;  \NXLO{2} of
      \wave{1}{S}{0}}\\\hline
    \hline\hline
    \makerow{\rule[-1.2ex]{0pt}{4ex}  {\#    at $Q^{-1}$}}{$2$}{$4$}{$5$}{$4$}\\\hline
    \makerow{\rule[-1.2ex]{0pt}{4ex} {\#     at $Q^{0}$}}{$+0$}{$+1$}{$+5$}{$+7$}\\\hline
    \makerow{\rule[-1.2ex]{0pt}{4ex}  {\#     at $Q^{1}$}}{$+7$}{$+8$}{0}{$+3$}\\\hline\hline
    \makerow{\rule[-1.2ex]{0pt}{4ex}total at $Q^{1}$}{$9$}{$13$}{$10$}{$14$}\\
    \hline
  \end{tabular}
  \caption{\label{tab:pc} Order $Q^n$ at which some LECs and the One- as well as
    Two-Pion-Exchange (OPE, TPE) enter in partial waves, for
    proposed power-countings in $\mathrm{NN}$  \ChiEFT~\cite{Saclay}. LECs
    of mixing angles are denoted \emph{e.g.}~by \wave{3}{SD}{1}. The bottom part
    summarises the number of LECs  at a given order. Not all  schemes
    have contributions at a given order, and some do not list all
    higher partial waves. While the information was collected with
    feedback from the respective authors, only I am to blame for
    errors. The results of Weinberg's pragmatic proposal have been shifted by
    $-1$ so that its potential  starts at order $Q^{-1}$, as mandated by the
    general arguments of
    eq.~\protect\eqref{eq:consistency}. %fig.~\protect\ref{fig:consistency}. 
  }
%\end{minipage}
%\twocolumn
\end{table*}
%\clearpage
%\end{strip}
%%%%%%%%%%%%%%%%%%%%%%%%%%%%%%%%%

%\clearpage
%%%%%%%%%%%%%%%%%%%%%%%%%%%%%%%
\section{The Test: Turning Cutoff Dependence into an Advantage}
\label{sec:test}

Assume we calculated an observable $\calO$ whose first nonzero contribution is
at order $n_0$ up to and including order $Q^n$ with $n\ge n_0$ in an EFT,
\emph{i.e.}~up to \NXLO{n-n_0} (next-to-next-to-\dots-leading order, with
$n-n_0$ occurrences of ``next'') relative to LO:
%//
%\begin{strip}
\begin{equation}
  \label{eq:observable}
\begin{split}
  \calO(k,\ptyp;&\Lambda;\LambdaEFT)=
  \sum\limits_{i=n_0}^n\left(\frac{k,\ptyp}{\LambdaEFT}\right)^i
  \calO_i(k,\ptyp;\LambdaEFT)\;\\& +\;\calC_n(\Lambda;k,\ptyp,\LambdaEFT)\left(\frac{k,\ptyp}{\LambdaEFT}\right)^{n+1-n_0}
\end{split}
\end{equation}
[Non-integer $n$ and non-integer steps from one order to the next will be
discussed in~\ref{para:extend}.] The notation
$\left(\frac{k,\ptyp}{\LambdaEFT}\right)$ for $Q$ indicates that numerators may depend
on combinations of both $k$ and $\ptyp$. If properly renormalised, effects
attributed to a regulator $\Lambda$ appear only at orders which are higher
than the last order $n$ which is known in full. Therefore, the sum $\calO$ may
depend on $\Lambda$ via higher-order cutoff artefacts which are subsumed into
the function $\calC_n$, but the renormalised order-$Q^i$ contribution $\calO_i$ to the observable $\calO$ does not. 

This series expansion of $\calO$ is based on a key assumption not only of EFTs
but of Physics in general: ``Naturalness''. To formulate this principle
rigorously is beyond the scope of this presentation. Sometimes, it is only
discussed as constraint on the LECs in an EFT, but the concept applies more
broadly. Paraphrasing van Kolck at the workshop out of which this volume
grew~\cite{vanKolckSaclay}, I define \emph{Weak Naturalness} as requiring that
higher-order terms in an expansion do generally not spoil the perturbative
series, \ie~$|\calO_{i}|>|\calO_{i+1}Q|$ in most cases; see
also~\cite{Hammer:2019poc}.

Weak naturalness is not quite as strong a condition as what many people
understand as ``Naturalness'', where all the coefficients shall be of ``order
$1$'', \ie~$l^n|\calO_i|\sim\calO(1)$, with the exponent $n$ of an appropriate
scale $l$ picked such that the result has dimension-less units and one needs
to agree what numbers shall be considered as ``of order $1$''. In order to
circumvent specifying a scale $l$, one can postulate that ``the coefficients
$\calO_i$ should all have about the same size'' or
``$|\calO_{i+1}|/|\calO_{i}|\sim1$ should be about one''. That, however, still
begs the question what ratios can be considered as ``order $1$'': what about
$1/4$th, $4$, or $10$?

Weak Naturalness, on the other hand, is tied to the size of the expansion
coefficient and can thus be defined more robustly. When $Q\approx10^{-20}$ as
in nuclear corrections from Quantum Gravity at the Planck scale, then ratios
of $|\calO_{i+1}|/|\calO_i|\approx10^{15}$ may appear prohibitively large, but
the contribution of the $(i+1)$st term in the series of the observable $\calO$
is still suppressed by $10^{-20+15}=10^{-5}$ against that of the $i$th term
and hence provides for all practical purposes a negligible correction. If, on
another hand, $Q\approx\frac{1}{4}$ as expected in \ChiEFT, then ratios of
$|\calO_{i+1}|/|\calO_i|\approx3$ or so are already precarious.

Naturalness for Quantum Field Theories was popularised by 't
Hooft~\cite{tHooft:1979rat}. It flows into the other fundamental EFT
assumption: Higher-order terms can reliably be estimated by Na\"ive
Dimensional Analysis, see \eg~\cite{NDA,NDA2,Griesshammer:2005ga}. Both
concepts have been implicit assumptions of the quantitative scientific method
since it was first employed.  Without these variants of Occam's
Razor~\cite{Occam}, one cannot rule out alternative explanations via
extraordinarily large higher-order corrections. Scientists just have to hope
that Nature is not malevolent (``Raffiniert ist der Herrgott, aber boshaft ist
er nicht.''~\cite{Einstein}).

Weak Naturalness also applies to the residual $\calC_n$.  While it may still
depend on $\LambdaEFT$, $k$ and $\ptyp$, it should be of natural size for all
$k,\ptyp<\LambdaEFT$, so that its contribution is parametrically suppressed by
$Q^{n+1}$ relative to the known terms of the series. Specifically, we require
that $\calO_n/\calC_n\gtrsim Q$ for (statistically speaking) the ``wide
variety of residuals and orders available'', with only ``a few'' exceptions.
If that were not the case, cutoff variations would regularly produce
corrections which are comparable in size to the regulator-independent terms
$\calO_i(k,\ptyp;\LambdaEFT)$. This would contradict the EFT assumption that
higher-order corrections are usually parametrically small.

A quantitative analysis of what constitutes a ``regular'', ``parametrically
small'' or ``exceptional'' case, or a ``wide variety'' and ``a few'' cases, or
under which circumstances a scale should be considered ``unnatural'', requires
a comprehensive and generally accepted, quantitative theory of (Weak)
Naturalness and Na\"ive Dimensional Analysis. That, however, does appear at
present not to be fully available~\cite{anonymous}. It is certainly well
beyond the scope of this presentation. Work in this direction will most likely
employ Bayesian statistical analysis, starting from reasonable expectations
clearly formulated as priors; see \eg~refs.~\cite{JPhysG,JPhysG2} and
references therein. Braaten and Hammer provide interesting probabilistic
interpretations of ``unnaturalness'' in a square-well-plus-van-der-Waals
setting in sect.~2.2 of ref.~\cite{Braaten:2004rn}.

With these qualifiers, we progress to explore the relative difference of
$\calO(k,\ptyp;\Lambda)$ in eq.~\eqref{eq:observable} at any two
cutoffs\footnote{This corrects an error in Ref.~\cite{Griesshammer:2015osb}
  and leads to a more nuanced presentation from here on.}:
\begin{align}
  \label{eq:master}
\begin{split}
  &{\frac{\calO(k,\ptyp;\Lambda_1)-\calO(k,\ptyp;\Lambda_2)}
  {\calO(k,\ptyp;\Lambda_1)}=
  \left(\frac{k,\ptyp}{\LambdaEFT}\right)^{n+1}}\\&
  \;\;\;\;\times\;\frac{\calC_n(\Lambda_1;k,\ptyp,\LambdaEFT)-
    \calC_n(\Lambda_2;k,\ptyp,\LambdaEFT)}
  {\calO(k,\ptyp;\Lambda_1;\LambdaEFT)}\\
\end{split}\\
  \label{eq:master2}
\begin{split}
&=\left(\frac{k,\ptyp}{\LambdaEFT}\right)^{n+1-n_0}
\frac{\calC_n(\Lambda_1;\dots)-
    \calC_n(\Lambda_2;\dots)}
  {\calO_{n_0}(k,\ptyp;\LambdaEFT)}\\&
\hspace*{5ex}\;\times\;\left[1+
    O\left(\frac{k,\ptyp}{\LambdaEFT}\right)
  \right]\;\;.
\end{split}
\end{align}
%\end{strip}
The last expression uses the expansion of $\calO$ to leading order and
demonstrates that the leading polynomial dependence is on
$\left(\frac{k,\ptyp}{\LambdaEFT}\right)^{n+1-n_0}$. Corrections are
suppressed at least by another power of $\frac{k,\ptyp}{\LambdaEFT}$ and carry
a mild dependence on the cutoff $\Lambda_1$; see also
note~\ref{para:assumptions}. 

For more insight, multiply eq.~\eqref{eq:master} by
$\Lambda_1/(\Lambda_1-\Lambda_2)$ and take $\Lambda_2\to\Lambda_1$:
\begin{equation}
  \label{eq:rge} \frac{\Lambda}{\calO}\;\frac{\dd\calO}{\dd\Lambda}=\frac{1}{\calO}
  \left(\frac{k,\ptyp}{\LambdaEFT}\right)^{n+1}
  \;\frac{\dd\calC_n(\Lambda)}{\dd\ln\Lambda}\;\;.
\end{equation}
This is the operator on the left of Wilson's Renormalisation Group Equation
for the observable $\calO$: $\frac{\dd\ln\calO}{\dd\ln\Lambda}=0$. Wilson's
equation is balanced by a zero on the right-hand side, guaranteeing that
observables are independent of the renormalisation scheme. Note that
eq.~\eqref{eq:rge} features a \emph{total} derivative: LECs in $\calO$ are
readjusted as $\Lambda$ changes.

In practise and in eq.~\eqref{eq:rge}, an EFT at finite order $n$ and with
finite cutoff tolerates cutoff artefacts which are parametrically small,
\emph{i.e.}~at least of order $n+1$. Therefore, the term on the right-hand
side is not necessarily zero, but only must be parametrically suppressed,
namely at least one order higher than the last retained order in $\calO$. This
also limits the rate of change in the residual $\calC_n$: Consistent with Weak
Naturalness, an observable is ``perturbatively renormalised'' when the
right-hand side of eq.~\eqref{eq:master} is smaller than any (or at least the
vast majority) of the terms on the left-hand side. To some, this condition
implies $\Lambda$ can only be varied in a range around $\LambdaEFT$; the
functional dependence on $k$ and $n$ is then still a quantitative
prediction. Equation~\eqref{eq:master} therefore reveals quantitative aspects
of the Renormalisation Group evolution and can be utilised to falsify claims
of consistency in an EFT.

One can now vary $k$ and $\ptyp$ to read off both the order $n$ to which the
calculation is complete and the breakdown scale $\LambdaEFT$. Subsequently,
one can reconstruct also the $k$-dependent expansion parameter $Q$ from
eq.~\eqref{eq:Q} (see also note~\ref{para:estimating}). All that is feasible if
the cutoff behaviour cannot be eliminated in its entirety,
\emph{i.e.}~$\calC_n(\Lambda_1)\ne\calC_n(\Lambda_2)$ for at least some cutoff
pairs, and if the residuals $\calC_n$ vary more slowly with $k$ and $\ptyp$
than with $\Lambda$; see also practicalities in notes~\ref{para:regulator}
and~\ref{para:cutoffs}.
The results of such a fit may certainly be inconclusive; see extended remarks
in sect.~\ref{sec:notesofnote}, in particular sect.~\ref{sec:misc}.
But if higher orders are \emph{not} parametrically suppressed and the exponent
comes out \emph{smaller} than the PC prediction $n+1-n_0$, then the EFT is
necessarily not properly renormalised. As will be discussed in
sect.~\ref{sec:principle} (in particular notes~\ref{para:necessary}
and~\ref{para:sampling}), an exponent $\ge n+1-n_0$ does neither suffice to
demonstrate consistency, nor does it establish failure.

In its generality, this equation may be of limited value since it relies on
dis-entangling behaviours in a multi-dimensional space which is spanned by
several expansion parameters $\frac{k}{\LambdaEFT}$ and
$\frac{\ptyp}{\LambdaEFT}$, the latter usually coming from a number of typical
low-energy scales, \emph{e.g.}~$\mpi,\gamma_{t,s},\dots$ in \ChiEFT. These can
be hard to vary in practise when data is available only at the physical point;
but see a later discussion in~\ref{para:extendwindow}.

For first steps, it may be more practical to consider eq.~\eqref{eq:master} at
fixed $\ptyp$ and vary the scattering momentum only in a ``window of
opportunity'' $\frac{\ptyp}{\LambdaEFT}\ll \frac{k}{\LambdaEFT}\ll1$ where all
scales but $\frac{k}{\LambdaEFT}$ can be neglected, at least in the first few
orders. The catch is that the denominator in eq.~\eqref{eq:master} also
contributes to the $k$-dependence, with its LO contribution
\begin{equation}
\label{eq:prescribe}
%\calO(k,\ptyp;\Lambda;\LambdaEFT) =
\left(\frac{k,\ptyp}{\LambdaEFT}\right)^{n_0}
\calO_{n_0}(k,\ptyp;\LambdaEFT) %+\dots
\end{equation}
and higher orders providing parametrically small corrections as indicated in
eq.~\eqref{eq:master2}. This does not mean that the exponent of $k$ itself is
$n+1-n_0$. Rather, it may be $n+1-\eta$, where $\eta\le n_0$ is the exponent
characterising the $k$-dependence of the observable at LO.

A simple example may be helpful to illustrate that point. From
eq.~\eqref{eq:consistency}, the amplitude $\calA\propto1/(k\cot\delta-\ii k)$
of $\mathrm{NN}$ scattering scales as $Q^{-1}$, \ie~$k\cot\delta\sim Q^1$
($n_0=1$). According to the Effective-Range
Expansion~\cite{ERE1,ERE2,ERE3,ERE4} in scattering momenta,
\begin{equation}
  k\cot\delta =-\gamma+\frac{r_0}{2}k^2+\calO(k^4)\;\;.
\end{equation}
The effective range $r_0$ sets the radius of convergence, $kr_0\ll1$, \emph{i.e.}~$r_0\sim1/\LambdaNoPion$. The
inverse system size (inverse scattering length) $\gamma$ pro\-vides a
``typical low-momentum scale'' and scales indeed as
$\gamma\equiv\ptyp\sim Q^1$ as predicted, but it involves no $k$-dependence,
$\gamma\sim (k/\LambdaNoPion)^0$, \emph{i.e.}~$\eta=0<n_0=1$. The denominator
of eq.~\eqref{eq:master} is thus $k$-independent at LO, and largely remains so
for $kr_0\ll1$ at NLO, albeit the NLO contribution itself is quadratic in
$k$. The variation of $k$ in eq.~\eqref{eq:master} will thus at \NXLO{n-n_0}
pick up a functional dependence on
$\left(\frac{k}{\LambdaEFT}\right)^{(n+1-\eta)=(n+1)}$, and not on
$\left(\frac{k}{\LambdaEFT}\right)^{(n+1-n_0)=n}$.

In addition, recall that the regions in which an observable is not analytic in
$(k,\ptyp)$ contain the actually physically most relevant information. If any
observable were just a Taylor expansion in $Q$, little could be learned from
it. Therefore, an observable is dominated by its non-analyticities because
these encode the Physics of relevance. In contradistinction, the residuals
$\calC_n$ are dominated by unphysical cutoff artefacts since they encode
Physics incorrectly captured a momenta larger than those at which the EFT is
applicable. These are thus well-parametrised by polynomials in $k,\ptyp$.

Therefore, instead of prescribing the $k$-dependence of the observable as
$k^{n_0}$ via eq.~\eqref{eq:prescribe}, it is more useful to have the LO
observable itself prescribe a LO dependence $\sim (k/\LambdaEFT)^\eta$ from
eq.~\eqref{eq:master}, where $\eta$ encodes that possible non-analyticity at
LO and might, in the worst case, depend on $k$. Each higher order adds then at
least one factor of $\frac{k}{\LambdaEFT}$. The LO exponent $\eta$ is not
re-calibrated at higher orders since it is bound to change only by
parametrically small amounts; see eq.~\eqref{eq:master2}. Otherwise,
higher-order effects would upset the ordering at the heart of power-counting
(\cf~Weak Naturalness and Na\"ive Dimensional Analysis). The exponent $\eta$ can
be non-integer (encoding non-analyticities), and must obey $\eta\le n_0$ for
an observables whose LO is of order $Q^{n_0}$.

Further notes on reasonable observable choices can be found in
sect.~\ref{sec:pickingobservables}.

Therefore, the following test emerges from eq.~\eqref{eq:masternew} for
variations in one variable $k\gg\ptyp$:
%\\
%\begin{strip}
\begin{equation}
  \label{eq:masternew}
\begin{split}
  \lefteqn{\frac{\calO(k,\ptyp;\Lambda_1)-\calO(k,\ptyp;\Lambda_2)}
  {\calO(k,\ptyp;\Lambda_1)}}\\
&\;\;\;\;\to\left(\frac{k}{\LambdaEFT}\right)^{n+1-\eta}
  \times\;f(\Lambda_1,\Lambda_2;k,\ptyp,\LambdaEFT)
\end{split}
\end{equation}
%\end{strip}
%\noindent
where $f(\Lambda_1,\Lambda_2;k,\ptyp,\LambdaEFT)$ is a slowly varying function
of $k$ and $\ptyp$. Here, $\eta\le n_0$ encodes the dependence of the
observable on $k$ for $\ptyp\ll k\ll \LambdaEFT$ and does not change from
order to order, but is determined already at LO. A discussion of the size of
this ``window of opportunity'', and how to possibly extend it, is provided in
notes~\ref{para:window} and~\ref{para:extendwindow}.

Equation~\eqref{eq:master} is rigorously true, but extracting slopes at large
$k$ as in eq.~\eqref{eq:masternew} needs additional assumptions, including a
choice of the ``window of opportunity'' and a robust but flexible algorithm to
extract slopes. All these can be cast into priors which lead to statistical
likelihoods with clearly stated underlying assumptions. It should not be a
surprise that techniques developed in Bayesian analysis will be most
helpful. Their application to EFTs is rapidly evolving; see
\emph{e.g.}~\cite{JPhysG,JPhysG2} and references therein.

This ends the discussion of eq.~\eqref{eq:master} in a limited space of
limited variations of $k$ at fixed $\ptyp$; varying different parameters is
discussed in note~\ref{para:choice}. Exploration of the whole parameter space
$(k,\ptyp)$ provides more unambiguous answers but would certainly not be
chosen for first studies. Indeed, the example discussed in the next section
will explore the one-parameter fit to $k$.

Equations \eqref{eq:master} and \eqref{eq:masternew} are formulated in terms
of renormalised quantities only and therefore hold for any regulator, but they
are most useful for cutoffs: Answers in nonperturbative EFTs are usually found
only numerically and for a \emph{cutoff regulator}, \emph{i.e.}~for a
regulator which explicitly suppresses high momenta $q\gtrsim\Lambda$ in loops.
It is this case which we use to our advantage from now on.

Cutoffs are of course only sensible if all loop momenta are sampled which lie
in the domain of validity of the EFT, \emph{i.e.}~if
$\Lambda\gtrsim\LambdaEFT$.  Only then can the coefficients $\calC_n$ be
expected to be of natural size relative to $\LambdaEFT$ (with the caveats
mentioned around eq.~\eqref{eq:consistency-n}). Remember that these
coefficients are those of observables and therefore should be renormalised
already. Except for this, no particular assumption is necessary as to the size
of $\Lambda$ relative to $\LambdaEFT$ in eqs.~\eqref{eq:master} and
\eqref{eq:masternew}.  In dimensional regularisation and some other analytic
regularisation schemes, on the other hand, renormalisation can be performed
exactly and no cutoff $\Lambda$ or residual regulator scale appears in
observables at all ($\calC_n\equiv0$ for all $n$). Equations \eqref{eq:master}
and \eqref{eq:masternew} are then an exact zero, with no information about $n$
and $\LambdaEFT$. But doubts about proper renormalisation of a calculation
which is analytic at each step do not arise, so the test is moot anyway.

Further comments and qualifiers about the test are postponed to
sect.~\ref{sec:notesofnote}. It contains a reiterated and augmented discussion
of assumptions, consequences and limitations (sect.~\ref{sec:principle}); a
discussion about which observables are most promising
(sect.~\ref{sec:pickingobservables}); and other, more general remarks
(sect.~\ref{sec:misc}).

%%%%%%%%%%%%%%%%%%%%%%%%%%%%%%%
\section{An Application: Confirming the Hierarchy of $3\mathrm{N}$ Interactions in \EFTNoPion}
\label{sec:application}

Before continuing the discussion of the parameters of the test in
sect.~\ref{sec:notesofnote}, consider its first application (to my knowledge):
the \wave{2}{S}{\frac{1}{2}} $\mathrm{Nd}$ wave in \EFTNoPion, where
$\LambdaNoPion\sim\mpi$ and $\ptyp\sim\gamma_{t,s}$. Electroweak effects are
not accounted for, \emph{i.e.}~$\mathrm{nd}$ and $\mathrm{pd}$ scattering are
identical. It is well-known that in this channel, the $3\mathrm{N}$
interaction without derivatives does not follow simplistic PC rules (``just
count momenta'') which predict $H_0$ at \NXLO{2} or
$\calO(Q^0)$~\cite{Bedaque:2002mn, Platter:2009gz, Hammer:2019poc}. Instead,
it is needed at LO to stabilise the system (Thomas collapse, Efimov effect);
its scaling, $H_0\sim Q^{-2}$, follows from eq.~\eqref{eq:consistency-n} for
$n=3$. If the first momentum-dependent $3\mathrm{N}$ interaction $k^2H_2$
follows the simplistic argument and scales as $Q^2$, then new LECs need to be
determined from $3\mathrm{N}$ data only at \NXLO{4}. Therefore, one could find
$2\mathrm{N}$ interaction strengths from few-N data with only one new
$3\mathrm{N}$ datum up to an accuracy of better than
$Q^4\approx\frac{1}{3^4}\approx1$\% at low momenta.  This is crucial for
example for hadronic flavour-conserving parity violation since it considerably
extends the number of targets and observables~\cite{Griesshammer:2010nd,
  Vanasse:2018buq}.

Based on the asymptotic off-shell amplitudes, Refs.~\cite{Bedaque:2002yg,
  improve3body} proposed that $H_2$ is only suppressed by $Q^2$ relative to
LO, \emph{i.e.}~that calculations at \NXLO{2} or on the $10$\%-level do already need
one additional $3\mathrm{N}$ datum as input. In Ref.~\cite{effrange}, this was
confirmed and extended to a general scheme to find the order at which any
given $3\mathrm{N}$ interaction starts contributing. The argument analyses
perturbations to the asymptotic form of the LO integral equation. It
%procedure 
is not immediately transparent, as witnessed by a subsequent claim by Platter
and Phillips that a $k$-dependent $3\mathrm{N}$ interaction enters not earlier
than \NXLO{3}~\cite{Platter:2006ev}. Upon closer inspection, that claim was
later refuted by Ji and Phillips~\cite{Ji:2012nj}.

\begin{figure}[!ht] 
  \includegraphics[width=\linewidth]
    {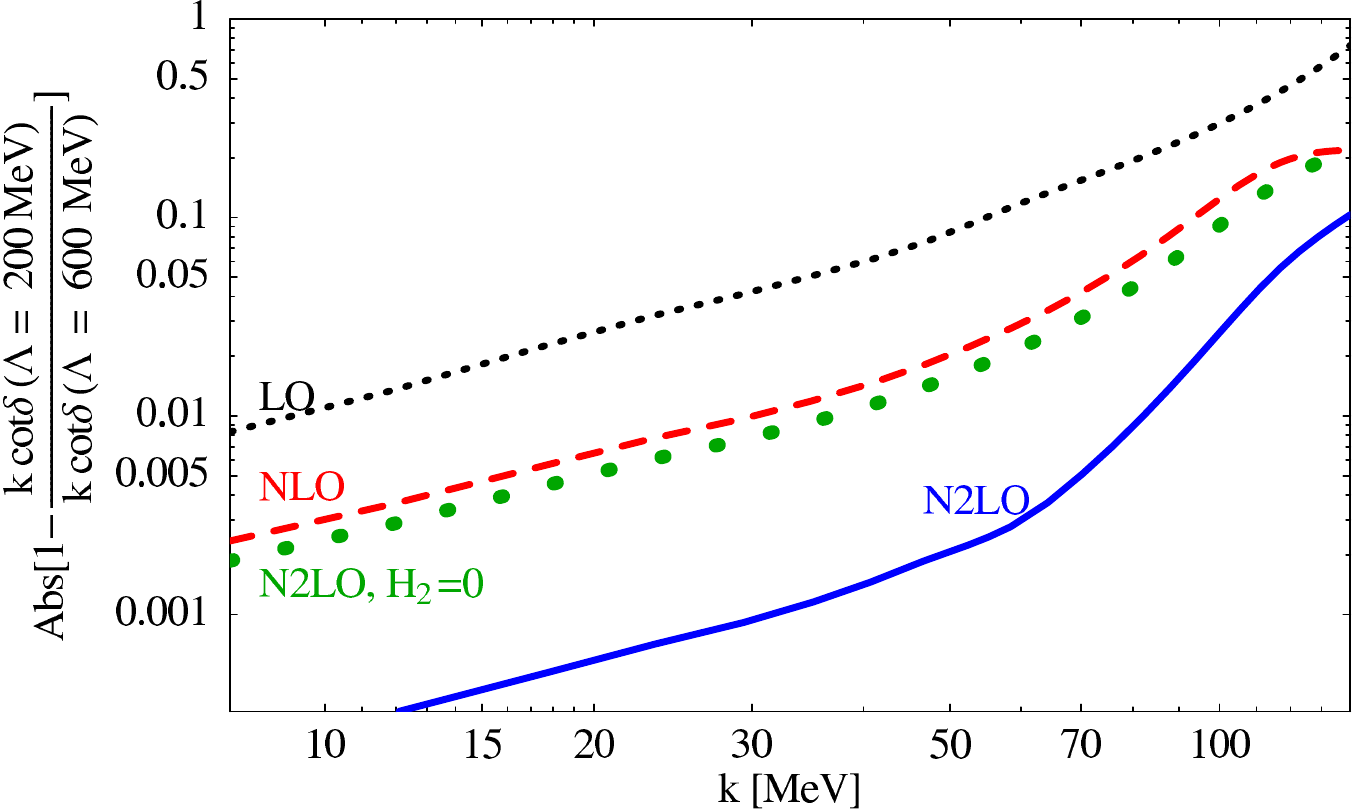}
    \caption{\label{fig:nonLepage} Double-logarithmic error plot for the
      \wave{2}{S}{\frac{1}{2}} wave of $\mathrm{Nd}$ scattering in \EFTNoPion;
      cf.~Refs.~\cite{Bedaque:2002yg, improve3body}. Black dotted line: LO;
      red dashed: NLO; blue solid: \NXLO{2} with $H_2\ne0$; green dotted:
      \NXLO{2} with $H_2\equiv0$.}
\end{figure}

Refs.~\cite{Bedaque:2002yg, improve3body} also supplied numerical evidence
from solutions of the Faddeev equations in momentum space with a step-function
cutoff: a double-logarithmic plot of eq.~\eqref{eq:master} for the inverse $K$
matrix, $\calO=k\cot\delta$ at $\Lambda_1=900\;\MeV$ and
$\Lambda_2=200\;\MeV$, both well above the breakdown scale
$\LambdaNoPion\approx\mpi$ of \EFTNoPion. A slight variant of that plot is
reproduced here as fig.~\ref{fig:nonLepage}, for a slightly smaller cutoff
$\Lambda_1=600\;\MeV$; fig.~\ref{fig:nonLepage2} contains the results for
$\Lambda_1=900\;\MeV$ and will be discussed later.

The cutoff dependence decreases order-by-order as expected when the theory is
perturbatively renormalised in the EFT sense. However, there is no decrease
from NLO (next-to-leading order) to \NXLO{2} when $H_2\equiv0$. That by itself
could be accidental -- after all, would one not expect better convergence with
one more parameter to tune? (No, see the subsequent discussions on the
influence of fitting parameters in sect.~\ref{sec:pickingobservables}.)

More informative is a look at the slopes in $\ln k$. Lines at different orders
are near-parallel for small $k$ because there are additional natural
low-energy scales $\ptyp$, namely the binding momenta of the deuteron
($\gamma_t\approx45\;\MeV$) and of the virtual singlet-S state
($\gamma_s\approx-8\;\MeV$). For $k\lesssim\gamma_{t,s}$,
eqs.~\eqref{eq:master} and \eqref{eq:masternew} are not very sensitive to $k$,
so all slopes should indeed be small and near-identical. However, in the
``window of opportunity'' $k\gg\gamma_{t,s}$ (but of course still
$k\ll\LambdaNoPion$, so that the EFT converges), they converge towards one
region. According to the discussion below eq.~\eqref{eq:masternew}, slopes in
that window are at order $Q^n$ given by $n+1-\eta$, with $\eta\le n_0$
determined by the LO-dependence on $k$.

Indeed, the fits of $n$ to the nearly straight lines in the momentum range
between $70\;\MeV>\gamma_{t,s}$ and $100$-to-$130\;\MeV\lesssim\LambdaNoPion$
compare well to the PC prediction when $H_2$ is added at
\NXLO{2}~\cite{improve3body}:
%\vspace{-0.5ex}
\begin{equation}
\label{eq:3Ntab}
\small
  \begin{tabular}{c@{\hq}|@{\hq\hq}c@{\hq\hq}c@{\hq\hq}c@{\hq\hq}|@{\hq\hq}c}
    slope& LO&NLO&\NXLO{2}&\NXLO{2} {without $H_2$}\\%[1ex]
    $n+1-\eta$&$n=n_0=2$&$n=3$&$n=4$&$n=4$\\
    \hline
    \rule{0ex}{2.5ex} slope fit &${\approx 1.9}$&{$2.9$}&{$4.8$} [\emph{sic!}]&${3.1}$\\%[1ex]
     prediction &${2}^\ast$&{$3$}&{$4$}&not renormalised
  \end{tabular} 
\end{equation}
The asterisk $^\ast$ serves as reminder that in the $3\mathrm{N}$ system,
$k\cot\delta\sim Q^2$ ($n_0=2$) from eq.~\eqref{eq:consistency-n}, but the
observed LO slope at large $k$ is $(n=n_0=2)+1-\eta\stackrel{!}{\approx}1.9$,
forcing $\eta=1$ as input into the following predictions. 

Without $H_2$ at \NXLO{2}, the slope does not improve from NLO. This is a
clear signal that the PC is inconsistent without a momentum-dependent $3\mathrm{N}$
interaction at \NXLO{2}: Its assumptions do not bear out in the functional
behaviour of this observable on $k$.  On the other hand, when $H_2$ is
included, the slope is markedly steeper than at NLO. The general agreement
between predicted and fitted slope is astounding, and actually quite stable
against variation of the fit range or of the two cutoffs $\Lambda_1$ and
$\Lambda_2$. Only the LO numbers are somewhat sensitive, and only to the upper
limit~\cite{improve3body}.

It may appear somewhat surprising that the slope increases by two units from
NLO to \NXLO{2} when one includes $H_2$. One would have expected the change
from each order to the next to be by only one unit. This may however stem from
the ``partially resummed formalism'' used at that time, which resums some
higher-order contributions. It may be worth revisiting this issue with
J.~Vanasse's method to determine higher-order corrections in ``strict
perturbation''~\cite{Vanasse:2013sda}; see also the extended
note~\ref{para:higher}. But we will see in the notes~\ref{para:assumptions} on
``\emph{Assumptions of the Expansion}'' that a fitted slope which is
\emph{larger} than predicted does \emph{not} invalidate the power counting --
the converse does.

Finally, the figure provides a rough value of
$\LambdaNoPion\approx[120\dots150]\;\MeV$ as the region where the fitted lines
of different orders coalesce. This is not in disagreement with the breakdown
scale expected of \EFTNoPion.

\begin{figure}[!h] 
\centering
  \includegraphics[width=0.8\linewidth]
    {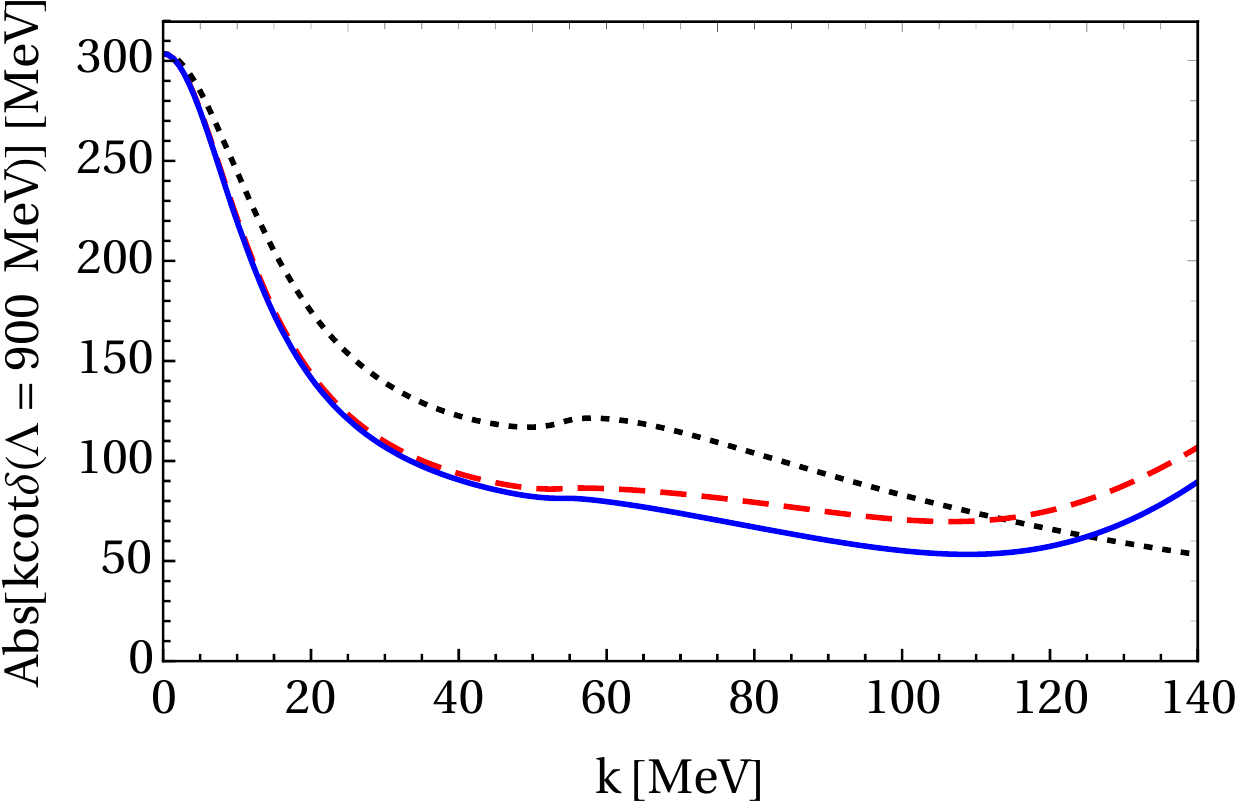}
    \caption{\label{fig:kcotdelta} The magnitude of $k\cot\delta$ at cutoff
      $\Lambda=900\;\MeV$ in the \wave{2}{S}{\frac{1}{2}} wave of
      $\mathrm{Nd}$ scattering in \EFTNoPion; cf.~Refs.~\cite{Bedaque:2002yg,
        improve3body}. Notation as in fig.~\ref{fig:nonLepage}.}
\end{figure}

For completeness, fig.~\ref{fig:kcotdelta} provides a plot of
$|k\cot\delta(\Lambda=900\;\MeV)|$ ($k\cot\delta$ is complex above the
$\mathrm{Nd}$ breakup, $k_{\mathrm{break}}\approx 52\;\MeV$).  It shows that
corrections from LO to NLO, and further on to \NXLO{2}, are indeed
parametrically small up to $k\lesssim140\;\MeV$. This provides information on
the magnitude of each contribution, while the error plot of
fig.~\ref{fig:nonLepage} provides information on the magnitude of the
\emph{variation} of each contribution with $\Lambda$. The extracted breakdown
scale of both agrees, as is to be expected.

%%%%%%%%%%%%%%%%%%%%%%%%%%%%%%%
\section{Notes of Note}
\label{sec:notesofnote}

With this example in mind, we conclude by considering assumptions, strengths,
extensions, features, caveats and limitations of such an analysis to assess
the consistency of a PC proposal, grouped by three topics: comments directly
relevant to the test; notes about the choice of observables, and concluding
remarks. With in each topic, arguments progress from general to specific.

\subsection{Matters of Principle}
\label{sec:principle}

Let us first discuss more details about the fundamental assumptions,
consequences and limitations of the procedure.

\paragraph{Extending the Expansion} \label{para:extend} In its original
formulation, eq.~\eqref{eq:master} may at first glance insinuate that each
order contains integer powers of the expansion parameter $Q$. However, the
order $n$ is not necessarily an integer, and the first omitted order is not
always $Q^{n+1}$, but more generally $Q^{n+\beta}$,
$\mathrm{Re}[\beta]>0$. For example, some \ChiEFT $\mathrm{NN}$ proposals in
table~\ref{tab:pc} proceed in half-integer steps. To replace $n+1\to n+\beta$
in eqs.~\eqref{eq:observable}, \eqref{eq:master}, \eqref{eq:masternew},
\eqref{eq:rge} -- and indeed throughout -- is straightforward. In \EFTNoPion,
the slope-fit in eq.~\eqref{eq:3Ntab} endorses that the $3\mathrm{N}$ PC
proceeds in integer steps.  Including non-analytic dependencies of the
residuals on $k$ or $\ptyp$ is also straightforward. For the remainder of the
presentation, all such replacements are implied, but we stick to the integer
case for convenience. One might also recall that arbitrarily small steps
$\beta\to0$ are not helpful unless the expansion parameter is extremely
small. In that case, $\lim_{\beta\to0}Q^\beta=1$ for fixed $Q$ implies that
there is no ordering of terms in the series by relative size.

\paragraph{Assumptions of the Expansion} \label{para:assumptions} It is also
appropriate to highlight and reiterate a few key premises. The assumptions on
the residual $\calC_n$ are endorsed if order $n$ and breakdown scale
$\LambdaEFT$ follow indeed the functional form of eqs.~\eqref{eq:master} or
its variants~\eqref{eq:masternew} and~\eqref{eq:rge}. Na\"ive Dimensional
Analysis sets the magnitude of $\calC_n$ to the scale of its
running~\cite{NDA, NDA2}. Its cutoff-dependence and other effects are
eventually absorbed into higher-order LECs, \emph{i.e.}~the cutoff dependence
of observables should generically decrease order-by-order -- even when no new
fit parameters/LECs are encountered (see also below).

We can actually be somewhat more specific about the condition that the
variation of the residual $\calC_n$ with respect to $\Lambda$ should be larger
than that for other parameters. Since $k,\ptyp\ll\Lambda_1,\Lambda_2$, the
numerator on the left-hand side of eq.~\eqref{eq:master} can be expanded as
%\\
%\begin{strip}
\begin{equation} 
\label{eq:residualexpansion}
\begin{split}
  %\frac{
\calC_n&(\Lambda_1;k,\ptyp,\LambdaEFT)-
    \calC_n(\Lambda_2;k,\ptyp,\LambdaEFT)%} 
%  {\calO(k,\ptyp;\Lambda_1;\LambdaEFT)}
\\
  &= c_0(\Lambda_1,\Lambda_2;\LambdaEFT)+
  c_1(\Lambda_1,\Lambda_2;\LambdaEFT)\;\frac{k,\ptyp}{\Lambda_1,\Lambda_2}+\dots
\end{split}
\end{equation}
% \end{strip}\noindent
If the first term dominates, then the dependence of eq.~\eqref{eq:master} on
$k$ and $\ptyp$ is indeed indicative of the order $Q^{n+1}$. If subsequent
terms dominate, the exponent of eq.~\eqref{eq:master} may be larger than $n+1$
-- but never smaller. Likewise, the slope of eq.~\eqref{eq:masternew} in the
``window of opportunity'' may be larger than $n+1-\eta$ (with $\eta\le n_0$)
-- but never smaller.

\paragraph{Necessary but Not Sufficient} \label{para:necessary} This last
argument shows that an exponent smaller than $n+1$ conclusively demonstrates
failure of the PC to be consistent. However, the criterion is necessary rather
than sufficient: Exponents $\ge n+1$ (slopes $\ge n+1-\eta$ in the ``window of
opportunity'') are proof neither of failure, nor of success. Indeed, a PC may
be inconsistent but the coefficient of the terms with exponent $<n+1$ may be
anomalously small, leading to a ``false negative'': The test does not reveal a
problem, but the power counting is still inconsistent. Only understanding the
limitations of the test allows one to avoid the danger of interpreting this as
a ``successful pass''.

\paragraph{Estimating the Expansion Parameter} \label{para:estimating} As
elaborated above, the test provides a direct prescription to find
$\LambdaEFT$, and thus an estimate of the momentum-dependent expansion
parameter $Q=\frac{k,\ptyp}{\LambdaEFT}$ as a function of $k$. But it also
allows for another practical way to assess $Q$: vary both cutoffs $\Lambda_1$
and $\Lambda_2$ over a wide range\footnote{Some claim that renormalisability
  requires that $\calO$ has a unique limit as $\Lambda\to\infty$.}. Ratios
between different orders estimate $Q(k,\ptyp)$, and hence residual theoretical
uncertainties as a function of $k$. This is of course only one of several ways
to assess $Q(k)$; within reason, the least optimistic and hence most
conservative of several methods should be picked. For example,
Ref.~\cite{Griesshammer:2011md} combined this with the convergence pattern of
the EFT series~\cite{Furnstahl:2015rha}.

\paragraph{Choice of Expansion Parameter} \label{para:choice} In
sect.~\ref{sec:application}, $k$ is varied while the other scales $\ptyp$ are
fixed, but any combination of the low-energy scales may serve as
variable(s). For example, scanning in the pion mass at fixed $k\ll m_\pi$ may
elucidate the $m_\pi$-dependence of some couplings. Recall that the chiral
limit $\mpi\to0$ is of obvious importance in \ChiEFT since its formal starting
point is chiral symmetry. This was indeed explored in Ref.~\cite{Beane:2001bc}
to demonstrate the inconsistency of Weinberg's pragmatic proposal. It can also
be of particular relevance to extrapolating lattice computations at
non-physical pion masses. Likewise, varying the anomalous scales
$\gamma_{t,s}$ may reveal information about how the EFT approaches the
unitarity limit $\gamma_{t,s}\to0$, whose importance for nuclear systems
recently has been emphasised; see \eg~\cite{Kievsky:2015dtk, Konig:2016utl,
  Konig:2016iny, Kolck:2017zzf, Kievsky:2018xsl, vanKolck:2019qea, Konig:2019xxk}. Here, I
will continue to concentrate on variations with $k$, but most issues transfer
straightforwardly to other variations.

\paragraph{Window of Opportunity} \label{para:window} As stressed around
eq.~\eqref{eq:master}, one can read off logarithmic slopes most easily in the
range $\ptyp\ll k\ll\LambdaEFT$. In the \EFTNoPion example above, that
``window of opportunity'' is narrow but appears to suffice:
$\LambdaNoPion/(\ptyp\sim\gamma_{t,s})\lesssim3$. In \ChiEFT with dynamical
$\Delta(1232)$ degrees of freedom, one would expect a wider range:
$\LambdaChi/(\ptyp\sim m_\pi)\gtrsim4$. Still, the one-pion exchange scale of
$\mathrm{NN}$ scattering, $\Lambda_\mathrm{NN}\approx300\;\MeV$, which was
mentioned in the Introduction, complicates the situation~\cite{Kaplan:1998we,
  Barford:2002je}.  If it is a low scale,
$\mpi<\Lambda_\mathrm{NN}\stackrel{?}{\ll}\LambdaChi$, the window of
opportunity could possibly be halved to
$\LambdaChi/(\ptyp\sim m_\pi,\Lambda_\mathrm{NN})\approx2$. In \ChiEFT without
explicit $\Delta(1232)$, $\Lambda_\mathrm{NN}$ is comparable to
$\LambdaChi(\slashed{\Delta})\approx300\;\MeV$, therefore counts as a high
scale, and one finds again a window of size
$\LambdaChi(\slashed{\Delta})/(\ptyp\sim m_\pi)\approx2$. Only further,
practical investigation can elucidate how big the window actually is in either
EFT.

One may of course also fit the variables $n$ and $\LambdaEFT$ in
eq.~\eqref{eq:master} to the numerical results below that window, but then one
needs to specify the scales $\ptyp$ and determine their contributions relative
to $k$. This could be achieved by independently varying $\ptyp$; see
note~\ref{para:choice} above. Alternatively, one can employ another trick
discussed now.

\paragraph{Extending the Window of Opportunity} \label{para:extendwindow} In
the physically accessible world, the size of that window is prescribed because
the scales $\ptyp\sim\mpi,\gamma_{t,s},\dots$ have fixed values. We can
however not only explore functional dependencies via variations of $\ptyp$
directly as in~\ref{para:choice}; we can simply extend the window in which the
slope of $k$ can be extracted by decreasing the scales $\ptyp$. While data is
available only at the physical point, an EFT power counting must remain
formally consistent when its underlying \emph{qualitative} assumptions are
still valid. In \ChiEFT, that is the interpretation of the pion as
quasi-Goldstone boson of chiral symmetry ($\mpi\ll\LambdaChi$), and the
existence of anomalously shallow binding scales in the $\mathrm{NN}$ system
($\gamma_{t,s}\ll\mpi$), which in turn is related to the importance of the
unitarity limit~\cite{Kievsky:2015dtk, Konig:2016utl, Konig:2016iny,
  Kolck:2017zzf, Kievsky:2018xsl, vanKolck:2019qea, Konig:2019xxk}. These
assumptions do not depend on the particular values of $\mpi$ or $\gamma_{t,s}$
off the physical point, so there is no need to explore how the parameters
$\gamma_{t,s}$ correlate with $\mpi$; that dependence was first studied in
refs.~\cite{Beane:2002xf, Epelbaum:2002gb}. Rather, one constructs multiverses
with different pion masses and $\gamma_{t,s}$, only for the purpose of
disentangling the multi-variate dependencies in eq.~\eqref{eq:master} and
enlarging the window of opportunity in which a slope in $k$ should emerge.

\paragraph{Choice of Regulator} \label{para:regulator} Residual cutoff
dependence emerges naturally in numerical computations. This test uses it as a
tool to check consistency, but how crucial are details of the regularisation
procedure? The example in sect.~\ref{sec:application} used a ``hard'' cutoff,
but $\LambdaEFT$ and the exponent $n$ do not depend on a specific
regulator. If the theory can be renormalised exactly, all residual regulator
dependence disappears by dimensional transmutation; cf.~\eqref{eq:rge}. It may
well be that the test is most decisive for regulators which are usually
disfavoured because they show significant cutoff artefacts (but in which the
residuals $\calC_n$ are of course still of natural size).

\begin{figure}[!htb]
  \centering \includegraphics[width=\linewidth]
  {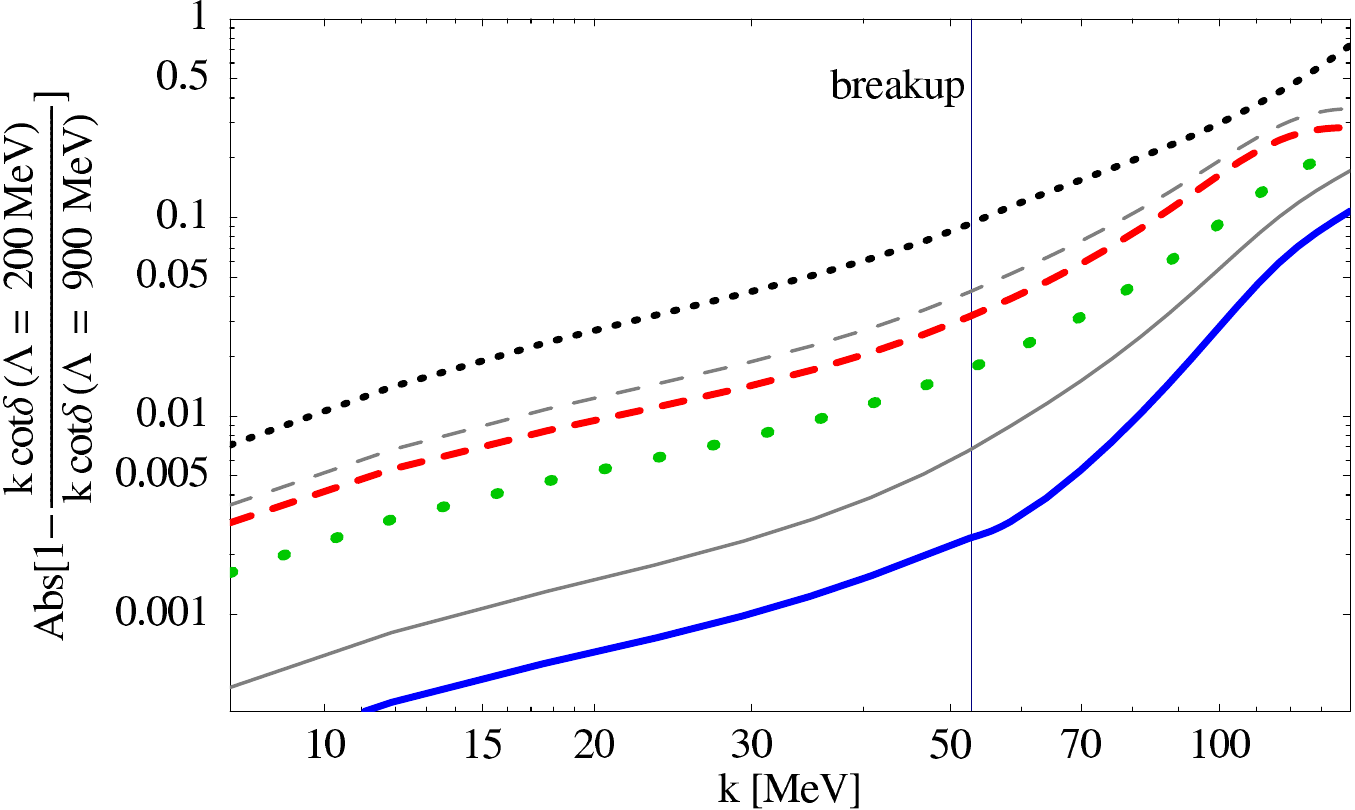}
  \caption{\label{fig:nonLepage2} \label{fig:nonLepage900MeV} \label{fig:EREvsZ}
    Thick coloured lines: Z-parametrisation of the $\mathrm{NN}$ amplitude as
    in fig.~\protect\ref{fig:nonLepage}, but for $\Lambda_1=900\;\MeV$, not
    $600\;\MeV$; thin gray lines: Bethe's Effective Range Parametrisation,
    whose LO is identical to that of Z-parametrisation; from
    Ref.~\protect\cite{improve3body}. The \NXLO{2} result for $H_2=0$ in
    Z-parametrisation is included as in fig.~\protect\ref{fig:nonLepage}
    (thick green dots). This provides an example that different choices for
    the cutoff $\Lambda_1$ lead to different curves but similar slopes
    (compare the thick coloured lines to fig.~\ref{fig:nonLepage}), and that
    this holds also when different parametrisations are used (gray
    vs.~coloured lines).  Overall notation as in fig.~\ref{fig:nonLepage}.}
\end{figure}

\paragraph{Choice of Cutoffs} \label{para:cutoffs} As just established, the
functional dependencies of eqs.~\eqref{eq:master} and \eqref{eq:rge} on $n$
and $\LambdaEFT$ do not depend on $\Lambda_1$ and $\Lambda_2$. While any two
cutoffs $\Lambda_1,\Lambda_2\gtrsim\LambdaEFT$ will do in principle, small
leverage may however lead to numerical artefacts. The larger
$\Lambda_2-\Lambda_1$, the clearer the signal should be. For our example,
fig.~\ref{fig:nonLepage900MeV} shows that an upper cutoff of $900\;\MeV$
instead of $600\;\MeV$ leads indeed to different curves but very similar
slopes, including for the \NXLO{2} result at $H_2=0$. Infinities, zeroes and
oscillations of $\calO$ with $k$ for any pair $\Lambda_1,\Lambda_2$ can lead
to problems (see note~\ref{para:accidental}) which are readily avoided by
choosing a cutoff pair such that $\calO(\Lambda_1)-\calO(\Lambda_2)>0$ for all
$k$. Even when one does not choose to take one of the cutoffs to
infinity\footnote{One could adhere to the philosophy that cutoffs and
  breakdown scales should be similar.}, a reasonable range of allowed cutoffs
exists. If $\Lambda_1\approx\Lambda_2$, one may of course directly consider
the numerical derivative of eq.~\eqref{eq:rge} -- over a range of cutoffs. [To
reiterate: exact cutoff independence $\calO(\Lambda_1)\equiv\calO(\Lambda_2)$
for any cutoff pair is not considered.]\footnote{Aside from the comments in
  the preceding two footnotes, I am a follower of the ``democratic principle''
  that any cutoff is equally legitimate and valid, as long as
  $\Lambda\gtrsim\LambdaEFT$.}

\paragraph{Decreasing Cutoff Dependence} \label{para:decreasing}
Equation~\eqref{eq:master} is a variant of the Renormalisation Group evolution
of $\calO$, eq.~\eqref{eq:rge}, which in turn quantifies the fundamental EFT
tenet that observables must become order-by-order less sensitive to loop
contributions beyond $\LambdaEFT$, the range of applicability. Cutoff
dependence in observables should therefore generically decrease from order to
order, irrespective whether or not LECs are fitted. This does not apply to the
$\calC_n$ themselves, but it does apply to the entire left-hand side of
eq.~\eqref{eq:master}. Refitting LECs may of course help to absorb some cutoff
dependence.
Indeed, no new LECs enter at NLO in the example above ($H_0$ is just
refitted), and the cutoff dependence decreases from LO to NLO. While it is
conceivable that the residual $\calC_n$ is sometimes somewhat larger than
Na\"ive Dimensional Analysis predicts, it should apply ``most of the time'',
statistically speaking and after appropriate Bayesian priors have been
declared -- see the comments below eq.~\eqref{eq:observable}.

Still, a specific regulator form may produce a very small residual cutoff
dependence at one order but a significantly larger one at a subsequent order,
$\calC_n(\Lambda_1)-\calC_n(\Lambda_2)<
\calC_{n+1}(\Lambda_1)-\calC_{n+1}(\Lambda_2)$
for some $n$.  This may for example occur if the regulator produces only
corrections with even powers of $\Lambda$ and the numerics preserves this
symmetry at least approximately (\emph{e.g.}~because
$\Lambda_1\approx\Lambda_2$, allowing for a perturbative expansion). If this
overwhelms the expansion in $Q$, $\calO$ may indeed systematically become more
dependent on $\Lambda$ between some orders, but not between all. Nonetheless,
one should not just see some qualitatively improved cutoff dependence with
increasing order, but one must see the quantitatively predicted slopes emerge
for many orders: that they must be $\ge n+1-\eta$ at \NXLO{n-n_0}, provides a
rigorous lower bound; see eqs.~\eqref{eq:masternew} and
\eqref{eq:residualexpansion}.

\paragraph{Constructing a PC by Trial-and-Error} \label{para:trial} This last
point provides an opportunity. If the cutoff dependence of a given observable
does not decrease consistently between subsequent orders, caution may be
advisable. For example, $\Lambda$-dependence may increase from one order to
the next, but then decrease markedly when another full order with a new LEC is
included. This could signal that this LEC cures cutoff dependence already at a
lower order -- and hence that the PC is inconsistent. One should then study
the convergence pattern as the LEC is promoted to a lower order such that the
cutoff dependence decreases always between subsequent orders. This may help to
construct a consistent PC by trial-and-error and iteration. Remember also that
after a LEC starts contributing at a certain order, it is re-adjusted at each
subsequent order to absorb both cutoff effects and still match its determining
datum.

\paragraph{Calculating Higher Orders} \label{para:higher} Traditionally,
observables beyond LO have been found by ``partially resumming''
contributions, \emph{i.e.}~the power-counted potential is iterated like in
Weinberg's original suggestion. Since corrections to the LO potential are
defined as parametrically small, they can be included in ``strict
perturbation''. This avoids two recurring problems. First, spurious deeply
bound states can be generated by iteration. This usually becomes the more
problematic, the higher the order or the larger the cutoff
$\Lambda$~\cite{improve3body, Vanasse:2013sda}. Second, partial resummation
often softens the (unphysical) ultraviolet behaviour of the amplitude, so that
fewer LECs appear to be necessary to cure residual cutoff dependence. A
striking example of this is found yet again in the $3\mathrm{N}$ system of
\EFTNoPion~\cite{Gabbiani:2001yh}. There, a careless resummation of
effective-range contributions at LO appears to eliminate the need for the
three-nucleon interaction which is central for the Efimov effect. That also
happens to lead to results which are not supported by data. A ``strictly
perturbative'' approach may provide clearer signals for the PC test than a
partially-resummed one.

%%%%%%%%%%%%%%%%%%%%%%
\subsection{Picking Observables}
\label{sec:pickingobservables}

The comments in this section discuss that not all observables are equally
suited for clear results of the proposed test, and provide criteria to
identify those that are more likely to be.

\paragraph{Isolating Dynamical Effects} \label{para:dynamical} While any
observable could be chosen, those which are free from kinematic or other
constraints (\emph{e.g.}~from symmetries) are preferred.
Consider the scattering amplitude $\calA_l$ in the $l$th partial wave (for
simplicity, assume no mixing). Since it is complex, one could choose
$\calO=|\calA_l|$. However, unitarity relates
$\calA_l=1/(k\cot\delta_l-\ii k)$ to the phase shift $\delta_l$. This
constraint dominates when $\delta_l$ is between about $\pi/4$ and $3\pi/4$ --
which affects much of the $\mathrm{NN}$ S-wave phase shifts. Even outside this
interval, the additional contribution to eqs.~\eqref{eq:master} and
\eqref{eq:masternew} is not sensitive to dynamics.
In addition, analyticity dictates that phase shifts approach zero like
$k^{2l+1}$ for $k\to0$ in the $l$th partial wave. Since both numerator and
denominator in eq.~\eqref{eq:master} are then zero, $\calO=\delta_l$ is
dominated by numerical uncertainties as $k\to0$. This may not be a problem if
the region in which the slopes are determined is far away, but only a closer
inspection could tell if that holds.  Likewise, it may be prudent to eliminate
phase-space factors of effects from cuts in decay constants, production cross
sections, etc, to arrive at ``smooth'' functions to analyse.

A sensible choice for single-channel scattering appears thus to be
$\calO=k^{2l+1}\cot\delta_l$: It is only constrained to be real below the
first inelasticity, and imaginary parts are usually small above it. One can
then choose to consider its magnitude, or real and imaginary parts
separately~\cite{Dai:2017ont}. Indeed, the S-wave example above kept track of
the imaginary part by plotting
\begin{equation}
  \left|1-\frac{k\cot\delta_0(\Lambda_2)}{k\cot\delta_0(\Lambda_1)}\right|\;\;.
\end{equation}
While factors of $k$ formally cancel, one should remember that the numerics of
calculating $\calA$ (\emph{i.e.}~$k\cot\delta_0$) is more benign when the
powers of $k$ are kept.

In the \EFTNoPion example above, examining
\begin{equation}
  1-\frac{|k\cot\delta_0(\Lambda_2)|}{|k\cot\delta_0(\Lambda_1)|}
\end{equation}
is disfavoured. In that form, there appears for each pair of cutoff values a
momentum $k_0$ in a range around $100\;\MeV$ inside the ``window of
opportunity'' where the results of both cutoffs appear to agree (``accidental
zero''; see also note~\ref{para:fitting1} below).

\paragraph{Partial-Wave Mixing} \label{para:mixing} In the $\mathrm{NN}$
system, two partial waves with total angular momentum $J$ mix. The
corresponding unconstrained observables in the Stapp-Ypsi\-lanti-Metro\-polis
(SYM or ``nuclear-bar'') parametrisation are
\begin{equation}
  k^{2\pm1-2J}\bar{\delta}_{J\pm1}\;\;\mbox{ and }
  \;\;k^{-(2J+1)}\bar{\epsilon}_J\;\;.
\end{equation}
In the Blatt-Biedenharn parametrisation, the same rules apply for the
eigenphases, but $k^{-2}\epsilon_J$ is the unconstrained variable for the
mixing angle; see \emph{e.g.}~\cite{deSwart:1995ui}. These choices do not suffer from
unitarity constraints (except for being real below the first inelasticity) and
can be used directly.

\paragraph{Dependence on Parameter Input} \label{para:input} The test's goal
is to resort minimally to empirical data, ideally only relying on the
existence of anomalously shallow scales in a theory whose LO is
non-perturbative, but not on their exact values.  Instead, the goal is to
focus on consistency of the EFT. To which degree can one achieve that? First,
consider processes in which $\calO(k)$ is a parameter-free prediction,
\emph{i.e.}~its LECs are all known from some other process(es). To what extent
does the procedure depend on that choice? In the example, the two-nucleon
interactions were determined to match the Z-parametrisation of
$\mathrm{NN}$-scattering (by fitting to the pole position and residue of the
scattering amplitude)~\cite{Phillips:1999hh}. Figure~\ref{fig:EREvsZ} shows
that results with Bethe's Effective-Range Parametrisation have a markedly
different rate of convergence, but the extracted slopes and $\LambdaNoPion$
agree very well~\cite{improve3body}. Note that LO is identical in both
parametrisations.

\begin{figure}[!h]\includegraphics[width=\linewidth]
    {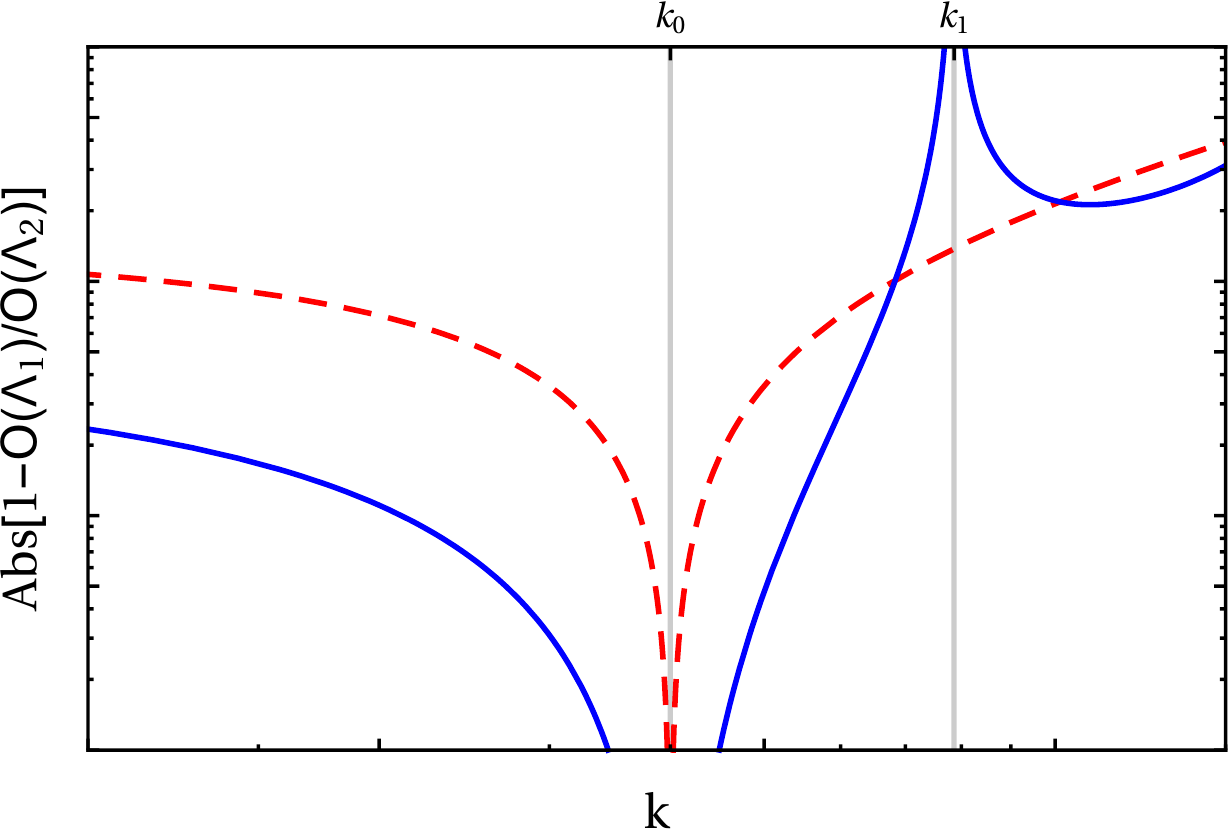}
  \caption{\label{fig:qualitative}
    Qualitative example of the impact of zeroes in
    $\calO(\Lambda_2)-\calO(\Lambda_1)$ (exact reproduction of datum at
    $k_0$), and in $\calO(\Lambda)$ (``accidental zero'' of $\calO(\Lambda_1)$
    at $k_1$).  Red dashed line: $n=1$; blue solid: $n=2$.}
\end{figure}

\paragraph{Accidental Zeroes and Infinities} \label{para:accidental} Some
observables may however show additional structures which should be
avoided. For example, the \wave{3}{P}{0} phase shift in $\mathrm{NN}$
scattering is zero at a lab energy of about $150\;\MeV$, so that the relative
deviation of $\calO=\delta_l$ in eq.~\eqref{eq:master} diverges. Likewise,
$\calO=k^{2l+1}\cot\delta_l$ diverges (approaches zero) at $\delta_l=0$
($\pi/2$), \emph{e.g.}~in the \wave{1}{S}{0} wave at $k\approx370\;\MeV$ and
\wave{3}{S}{1} wave at $k\approx90\;\MeV$ and
$400\;\MeV$~\cite{Epelbaum:2014efa}. As the qualitative plot in
fig.~\ref{fig:qualitative} shows, the corresponding spikes may make it more
difficult to determine slopes.

\begin{figure}[!htb]
  \includegraphics[width=\linewidth]{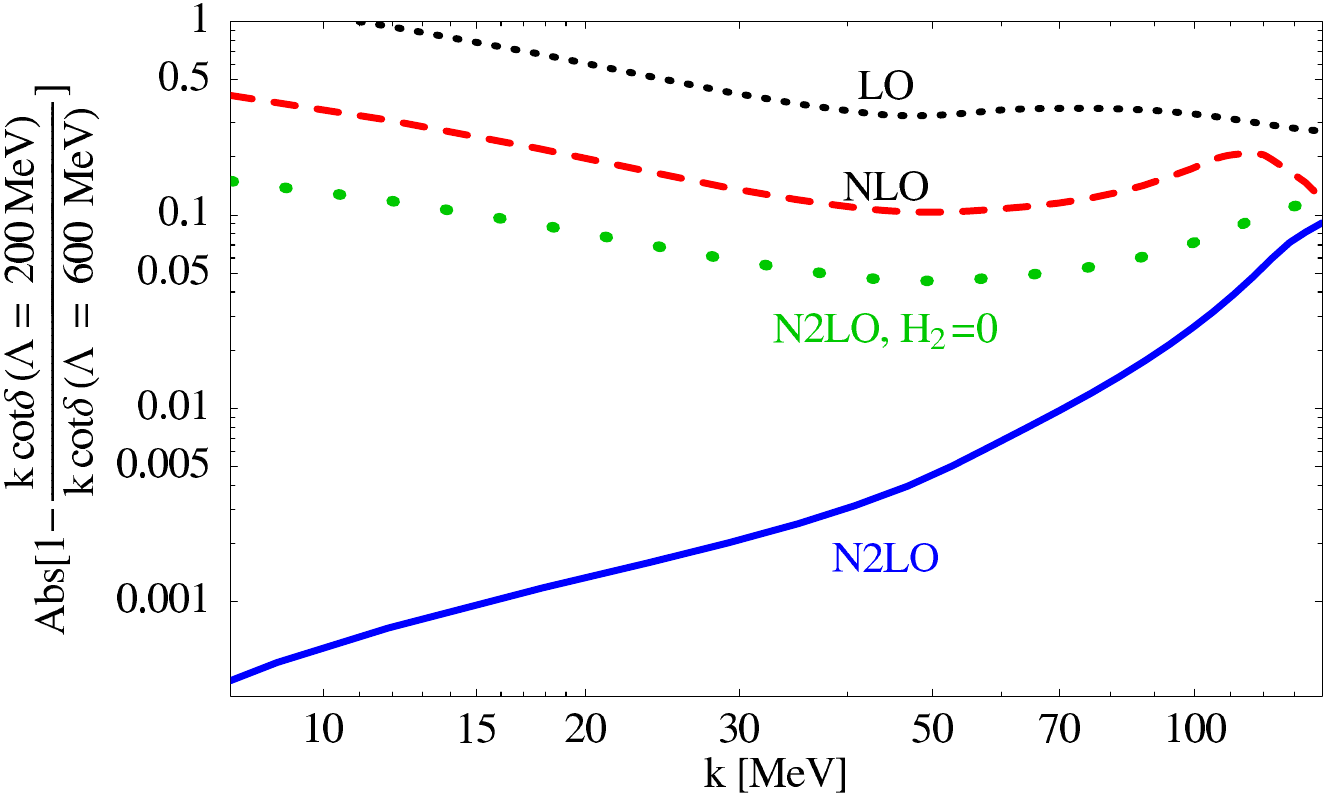}
  \caption{\label{fig:toB3} Test when
    the leading $3\mathrm{N}$ interaction is determined not by the $\mathrm{Nd}$ scattering
    length as in fig.~\protect\ref{fig:nonLepage}, but by the position of the
    triton pole. The \NXLO{2} fit uses again the $\mathrm{Nd}$ scattering length and
    triton binding energy.}
\end{figure}

\paragraph{Fitting to a Point} \label{para:fitting1} Such a ``zero'' in
eq.~\eqref{eq:master} is actually induced intentionally when the observable
contains a LEC that is determined in the channel in which one tests the PC. If
the observable is tuned to exactly reproduce a certain value at some point
$(k_0,\ptyp)$, then $\calO(k_0;\Lambda_1)-\calO(k_0;\Lambda_2)=0$ -- with all
the problems mentioned just now. Obviously, one should choose the fit point to
be outside the ``window of opportunity''. In the example of
sect.~\ref{sec:application}, the strength of the $3\mathrm{N}$ interaction
$H_0$ without derivatives was fixed at each order to the $\mathrm{Nd}$
scattering length, \emph{i.e.}~using $k=0$ as fit point. That is far away from
the ``window of opportunity''. At \NXLO{2}, the momentum-dependent
$3\mathrm{N}$ interaction $H_2$ was in addition determined from the triton
binding energy $B_3=8.48\;\MeV$, \emph{i.e.}~the pole in the amplitude is
fixed to $k_0=\sqrt{-4M B_3/3}\approx100\;\ii\;\MeV$. If one chooses this fit
point for $H_0$ at LO and NLO, instead of $k_0=0$, the pattern of the slopes
is wiped out; see fig.~\ref{fig:toB3}. It appears that fitting only at $k_0$
introduces a new low-energy scale $\ptyp$ and leaves no window
$\LambdaNoPion\gg k\gg|k_0|\approx100\;\MeV$, while the \NXLO{2} fit at both
$k=0$ and $k_0$ does not suffer this limitation.

\paragraph{Fitting in a Region} \label{para:fitting2} The issue is less
transparent when the LEC is not determined by exactly reproducing some data,
but by fitting over a whole region in $k$. That is the typical case in
$\mathrm{NN}$ scattering (see \emph{e.g.}~Ref.~\cite{Epelbaum:2014efa}), and
in $\overline{\mathrm{N}}\mathrm{N}$ scattering~\cite{Dai:2017ont}. The
deviation of the fitted result from data is more regular at any given cutoff
$\Lambda$ than when it is exactly zero at $k_0$. A pronounced spike is
therefore replaced by a more uniformly, but also more stealthily, constrained
behaviour inside the fit region. The comparison between two cutoffs in
eqs.~\eqref{eq:master} and \eqref{eq:masternew} is thus also more uniform as a
function of $k$. Since cutoff variations can now be balanced by adjusting
LECs, the coefficients $\calC_n$ are artificially small in that r\'egime. One
still expects the cutoff dependence to decrease order-by-order, but the
characteristic slopes are harder to see since the observable is constrained by
the fit. Just like in the neighbourhood of a fit point, an observable will
first have to shed the fit constraints outside the fit region for pronounced
slopes to emerge.

Such a fit region must of course be inside the applicability range of the
EFT. Traditional fits do not take into account that the systematic
uncertainties of an EFT increase with $k$, but assign a $k$-independent
uncertainty weight. Equation~\eqref{eq:observable} suggests that this is
justified for $k\lesssim\ptyp$ since the error varies only mildly. In that
case, one can speculate that the impact on the slopes at higher $k$ is not too
big. This limits a reasonable fit region to $k\lesssim\gamma_{t,s}$ in
\EFTNoPion; and to $k\lesssim\mpi$ in \ChiEFT. In addition, one expects
clearer signals if the same fit region is used at each order. It is difficult
to see how slopes can clearly be identified when the fit region extends far
towards $\LambdaEFT$. Practical considerations, like insufficient or
low-quality data at low momenta, may well override this choice. But there is a
way around, as discussed now.

\paragraph{Fitting to Pseudo-Data} \label{para:pseudodata} As a recourse and
in order to assess the impact of a fit region on the slopes, one may create an
artificial, ``exact datum'' $\calO_0(k_0)$ at very low $k\to0$ which agrees
with low-energy data (\emph{e.g.}~a scattering length, effective range, etc);
and then assess the dependence of the slope on reasonable variations of
$\calO_0(k_0)$. The goal is then not to find good agreement with actual data
at higher energies, but to test the convergence pattern. An EFT will not
become inconsistent just because some particular datum is shifted by some
small amount, or because the error bar on a datum is substantially
reduced. For example, the power-counting for $\mathrm{NN}$ amplitudes in
eq.~\eqref{eq:consistency} only uses that there \emph{is} a shallow real or
virtual bound state, not its exact location.

\paragraph{Summary: Choice of Observable} Ideal candidates for $\calO$ are
positive-definite observables which are not subject to unitarity and other
constraints, and which are nonzero and finite over a wide range in $k$ and
$\Lambda$, including the r\'egime $k\gtrsim\ptyp$ where one hopes to determine
the slope. EFT parameters/LECs should be determined at very low $k$. A good
signal may need some creativity. The choices $\calO=k^{2l+1}\cot\delta_l$,
$k^{2\pm1-2J}\bar{\delta}_{J\pm1}$ and $k^{-(2J+1)}\bar{\epsilon}_J$, with
Effective-Range parameters determining unknowns, appear suitable in most
scattering cases. 

%%%%%%%%%%%%%%%%%
\subsection{Miscellaneous Notes}
\label{sec:misc}

Finally, the following notes wrap up a variety of issues.

\paragraph{Consistency Assessment vs.~``Lepage Plots'' and Other Data-Driven
  Approaches} \label{para:lepage} To put this test into a broader context, one
should note that double-log\-arithmic convergence plots are not
unfamiliar. Lepage compared to data in order to quantify how accurately the
EFT reproduces experimental information~\cite{Lepage:1997cs}. This triggered a
series of influential studies of differences between approximations and
``exact results'' in toy-models, see \emph{e.g.}~\cite{Steele:1998un,
  Steele:1998zc, Kaplan:1999qa}. More recently, Birse and collaborators
perused similar techniques, after removing the strong influence of long-range
Physics (One- and Two-Pion Exchange) from empirical phase shifts in a modified
Effective Range Expansion. This allows a more detailed study of the residual
short-distance interactions under the assumption that long-dis\-tance Physics
is generally agreed upon to be understood sufficiently
well~\cite{Birse:2007sx, Birse:2010jr, Ipson:2010ah}. As table~\ref{tab:pc}
shows, that is true for one-pion exchange, but not for two-pion exchange. Both
of these alternative approaches rely on a high-quality, data-based partial
wave analysis.

The test advocated here emphasises somewhat different aspects. It tries to
minimise dependence on data, depending only on empirical input which is mostly
qualitative: the existence of anomalously small scales, and of some
low-momentum window in which a few data can be used to determine LECs. It then
aims to answer complementing questions: Does the output match the assumptions?
Is the theory consistent, and consistently renormalised in both its long- and
short-range aspects?  Recall that an EFT may converge by itself, but not to
data, if some dynamical degrees of freedom are incorrect or missing.
For example, a \ChiEFT without dynamical $\Delta(1232)$ at $k\approx300\;\MeV$
cannot reproduce Delta resonance properties -- but it may well be consistent.

In other words, an EFT may be consistent by itself, but not consistent with
Nature.

\paragraph{Insensitivity to Some LECs} \label{para:LECs} This procedure can
only help determine if a LEC is correctly accounted for when it is needed to
absorb residual cutoff dependence. Equation~\eqref{eq:rge} then determines its
running, and its initial condition is fixed by some input, for example data or
results of a more fundamental theory. Some LECs do however start contributing
just because of their natural size, and not to renormalise that order. For
example, the magnetic moment of the nucleon enters the one-baryon Lagrangean
of \ChiEFT at NLO, albeit it is not needed to renormalise loops. Similarly,
the contribution of a LEC to a particular observable may be unnaturally small
(or even zero).

\paragraph{Numerics}\label{para:numerics}  The analysis can be numerically
indecisive. We would trust results only if $n$ and $\LambdaEFT$ can be
determined quite robustly in a reasonably wide range to cutoffs (and,
possibly, cutoff forms), parameter sets and fit-windows. None of this
provides, however, sufficient excuse not to report results.

\paragraph{Sampling Tests} \label{para:sampling} Finding that the exponent at
each order $Q^n$ is not smaller than $n+1-n_0$ is necessary but not sufficient
for a consistent PC. We saw that fine-tuning, particular choices of regulator
forms and observables, and anomalously small coefficients are some reasons
which may hide signals of exponents $<n+1-n_0$ which violate the PC
assumptions. If exponents are always $\ge n+1-n_0$ for a variety of
independent observables, regulators etc., that may increase confidence in PC
consistency -- but cannot prove it. The same statement holds when the exponent
$n+1-n_0$ is substituted by the slopes $n+1-\eta$ in the ``window of
opportunity''.

 That is why this is a falsification test.

%%%%%%%%%%%%%%%%%
\subsection{Outlook} 

Unnaturally small scales provide significant challenges to formulating a
power-counting scheme in EFTs when analytic results cannot be obtained.  This
presentation advocated a quantitative and pragmatic test which can falsify a
proposed scheme, and which may elucidate the power-counting issues which
plague \ChiEFT. It provides a necessary but not sufficient consistency
criterion.

In its simplest and, for now, only tested variant, a ``window of opportunity''
is necessary, in which all low scales but one momentum scale $k$ can be
neglected, but in which the EFT still converges, $k\ll\LambdaEFT$. This may be
problematic in \ChiEFT because it is not quite clear what role is played by
$\Lambda_\mathrm{NN}\approx300\;\MeV$, the strength-scale of the $\mathrm{NN}$
potential; see discussion in~\ref{para:window}.

The \ChiEFT power-counting proposals differ most starkly in the attractive
triplet partial waves of $\mathrm{NN}$ scattering since they reflect different
philosophies on how to treat the non-selfadjoint, attractive $1/r^3$ potential
at short distances which appears at leading order; see table~\ref{tab:pc}. It
would therefore be interesting to see this test applied to the \wave{3}{P}{0}
wave and to the \wave{3}{P}{2}-\wave{3}{F}{2} system; and work is indeed under
way~\cite{Yang:2019hkn, hgrieYang}. In addition, one should explore whether
one can merge the present approach with the work of Birse and
collaborators~\cite{Birse:2007sx, Birse:2010jr, Ipson:2010ah}. As one referee
pointed out, it may be possible to analyse quasi-data generated at different
EFT orders using the modified Effective Range Expansion to more clearly
isolate short-distance effects.

The test proposed here is not necessarily a silver bullet to endorse or reject
a particular counting since its results may in the worst case be
inconclusive. It is thus neither more nor less than one more arrow in the
quiver to test EFTs. But that implies it is still worth a try.

%%%%%%%%%%%%%%%%%%%%%%%%%
\section*{Acknowledgements} 

I cordially thank the organisers of the workshop \textsc{The\linebreak Tower
  of Effective (Field) Theories and the\linebreak Emergence of Nuclear
  Phenomena (EFT and Philosophy of Science)} at CEA/SPhN Saclay in 2017 for
making a teenager's dream come true to discuss with Philosophers in France,
and all participants for the profound insight they shared, as well as for
enlightening and entertaining discussions.  These notes grew out of the
inspirational and intense discourses at the workshops \textsc{Nuclear Forces
  from Effective Field Theory} at\linebreak CEA/SPhN Saclay in 2013, \textsc{Bound
  States and Resonances in Effective Field Theories and Lattice QCD
  Calculations} in Benasque (Spain) in 2014, \textsc{Chiral Dynamics 2015} in
Pisa (Italy), \textsc{EMMI Rapid Reaction Task Force ER15-02: Systematic
  Treatment of the Coulomb Interaction in Few-Body Systems} at Darmstadt
(Germany) in 2016, and \textsc{New Ideas in Constraining Nuclear Forces} at
the ECT* in Trento (Italy) in 2018. I am most grateful to all their organisers
and participants. Since 2013, exchanges with M.\ C.\ Birse, B.\ Demissie, A.\
Ekstr\"om, E.\ Epelbaum, C.\ Forssen, R.\ J.\ Furnstahl, J.\ Holt, B.\ Long,
M.\ Pavon Valderrama, D.\ R.\ Phillips, M.\ J.\ Savage, I.\ Tews, R.\ G.\ E.\
Timmermans, U.\ van Kolck and Ch.-J.\ Yang allowed me to develop these ideas
into a sharper analysis tool. M.\ J.\ Birse, B.\ Demissie, E.\ Epelbaum, D.\
R.\ Phillips and Ch.-J.\ Yang suggested important improvements to this
script. I am especially indebted to ceaseless insistence on clarity by many
emerging researchers, and by both referees. Finally, my colleagues may forgive
mistakes and omissions in referencing work and historical precedents, and
graciously continue to point out necessary corrections.
This work was supported in part by the US Department of Energy under contract
DE-SC0015393, and by The George Washington University: by the Dean's Research
Chair programme and an Enhanced Faculty Travel Award of the Columbian College
of Arts and Sciences; and by the Office of the Vice President for Research and
the Dean of the Columbian College of Arts and Sciences; and was conducted in
part in GW's Campus in the Closet.

%%%%%%%%%%%%%%%%%%%%%%%%%
\section*{Data Availability Statement}

The preciously few data underlying this work are available in full upon request
from the author.

\end{document}